\documentclass[twocolumn,10pt,letterpaper]{article}
\usepackage[top=2cm, bottom=2.5cm, left=1.8cm, right=1.8cm]{geometry}
\usepackage{times}  %
\usepackage{helvet}  %
\usepackage{courier}  %
\usepackage{graphicx}  %
\usepackage{setspace}  %

\usepackage[small,bf]{caption}
\usepackage{subcaption}
\usepackage{amssymb,amsmath}

\usepackage{color}
\usepackage{ifthen}
\usepackage[T1]{fontenc}
\usepackage{newclude}
\usepackage{epstopdf}
\usepackage{xspace}
\usepackage{graphicx}
\usepackage{tikz}
\usepackage{pgfplots, pgfplotstable}
\usepgfplotslibrary{statistics}
\usepackage{xspace}

\usepackage[hang,flushmargin,stable]{footmisc}

\usepackage[hyphens]{url}

\makeatletter
\def\url@foostyle{%
  \@ifundefined{selectfont}{\def\UrlFont{\rm}}{\def\UrlFont{\rmfamily}}}
\makeatother
\usepackage[hang,flushmargin]{footmisc}
\newcommand{\descr}[1]{\smallskip\noindent\textbf{#1}}

\usepackage[compact]{titlesec}
\titlespacing*{\section}{0pt}{*3}{4.5pt} 
\titlespacing{\subsection}{0pt}{*3}{4pt}
\titlespacing{\subsubsection}{0pt}{*1.5}{3pt}

\let\OLDthebibliography\thebibliography
\renewcommand\thebibliography[1]{
  \OLDthebibliography{#1}
  \setlength{\parskip}{0pt}
  \setlength{\itemsep}{0pt plus 0.3ex}
}

\newcommand{\dsb}{{\sf{\small /b/}}\xspace}
\newcommand{\dspol}{{\sf{\small /pol/}}\xspace}
\newcommand{\dspolTTT}{/pol/\xspace}
\newcommand{\dspolCCC}{{{/pol/}}\xspace}
\newcommand{\dssp}{{\sf{\small /sp/}}\xspace}
\newcommand{\dsspCCC}{{{/sp/}}\xspace}
\newcommand{\dsint}{{\sf{\small /int/}}\xspace}
\newcommand{\dsintCCC}{{{/int/}}\xspace}

\newcommand{\hasht}[1]{{\sf{\small \##1}}\xspace}

\definecolor{darkred}{RGB}{153,0,0}
\definecolor{darkblue}{RGB}{0,0,119}
\hypersetup{colorlinks=true, linkcolor = darkred, citecolor = darkred, urlcolor = darkblue}

\makeatletter
\def\@copyrightspace{\relax}
\makeatother
\makeatletter
\def\@copyrightspace{\relax}
\makeatother
\begin{document}  

\sloppy

\title{\bf Kek, Cucks, and God Emperor Trump: A Measurement Study of 4chan's Politically Incorrect Forum and Its Effects on the Web\thanks{A shorter version of this paper appears in the Proceedings of the 11th International AAAI Conference on Web and Social Media (ICWSM'17). Please cite the ICWSM'17 paper. Corresponding author: blackburn@uab.edu.}}

\author{Gabriel Emile Hine$^\ddagger$,
Jeremiah Onaolapo$^\dagger$,
Emiliano De Cristofaro$^\dagger$,
Nicolas Kourtellis$^\sharp$,\\
Ilias Leontiadis$^\sharp$,
Riginos Samaras$^\star$,
Gianluca Stringhini$^\dagger$,
Jeremy Blackburn$^\sharp$\\[1ex]
{\normalsize $^\ddagger$Roma Tre University \hspace{0.2cm}
$^\dagger$University College London \hspace{0.2cm}
$^\sharp$Telefonica Research \hspace{0.2cm}
$^\star$Cyprus University of Technology}\\[-0.5ex]
{\normalsize gabriel.hine@uniroma3.it, \{j.onaolapo,e.decristofaro,g.stringhini\}@cs.ucl.ac.uk}, \\[-0.5ex]
{\normalsize  \{nicolas.kourtellis,ilias.leontiadis,jeremy.blackburn\}@telefonica.com, ri.samaras@edu.cut.ac.cy}\\}

\date{} 
\maketitle

\begin{abstract}
The discussion-board site 4chan has been part of the Internet's dark underbelly since its inception, and recent political events have put it increasingly in the spotlight. In particular, /pol/, the ``Politically Incorrect'' board, has been a central figure in the outlandish 2016 US election season, as it has often been linked to the alt-right movement and its rhetoric of hate and racism. However, 4chan remains relatively unstudied by the scientific community: little is known about its user base, the content it generates, and how it affects other parts of the Web. In this paper, we start addressing this gap by analyzing /pol/ along several axes, using a dataset of over 8M posts we collected over two and a half months. First, we perform a general characterization, showing that /pol/ users are well distributed around the world and that 4chan's unique features encourage fresh discussions. We also analyze content, finding, for instance, that YouTube links and hate speech are predominant on /pol/. Overall, our analysis not only provides the first measurement study of /pol/, but also insight into online harassment and hate speech trends in social media.
\end{abstract}

\section{Introduction}\label{sec:intro}

The Web has become an increasingly impactful source for new ``culture''~\cite{culture},
producing novel jargon, new celebrities, and disruptive social phenomena.
At the same time,
serious threats have also materialized, including the increase in hate speech and abusive behavior~\cite{blackburn2014stfunoob,www2016yahoopaper}. In a way, the Internet's global communication capabilities, as well as the platforms built on top of them, often enable previously isolated, and possibly ostracized, members of fringe political groups and ideologies to gather, converse, organize, as well as execute and spread their agenda~\cite{trolls}.

Over the past decade, \url{4chan.org} has emerged as one of the most impactful generators of online culture.
Created in 2003 by Christopher Poole (aka `moot'), and
acquired by Hiroyuki Nishimura in 2015, 4chan is an imageboard site, built around a typical discussion bulletin-board model.
An ``original poster'' (OP) creates a new thread by making a post, with a single image attached, to a board with a particular interest focus.
Other users can reply, with or without images, and add references to previous posts, quote text, etc.
Its key features include anonymity, as no identity is associated with posts,
and ephemerality, i.e., threads are periodically pruned~\cite{bernstein20114chan}.
4chan is a highly influential ecosystem: it gave birth not only to significant chunks of Internet culture and memes,\footnote{For readers unfamiliar with memes, we suggest a review of the documentary available at \url{https://www.youtube.com/watch?v=dQw4w9WgXcQ}.} but also provided a highly visible platform to movements like {\em Anonymous} and the {\em alt-right} ideology.
Although it has also led to positive actions (e.g., catching animal abusers), it is generally considered one of the darkest corners of the Internet, filled with hate speech, pornography, trolling, and even murder confessions~\cite{murder}. 4chan also often acts as a platform for coordinating denial of service attacks~\cite{ddos} and aggression on other sites~\cite{tumblr}.
However, despite its influence and increased media attention~\cite{wired,fortune}, 4chan remains largely unstudied, which motivates the need for systematic analyses of its ecosystem.

In this paper, we start addressing this gap, presenting a longitudinal study of one sub-community, namely, \dspol, the ``Politically Incorrect'' board.
To some extent, \dspol is considered a containment board, allowing generally distasteful content -- even by 4chan standards -- to be discussed without disturbing the operations of other boards, with
many of its posters subscribing to the alt-right and exhibiting characteristics of xenophobia, social conservatism, racism, and, generally speaking, hate.
We present a multi-faceted, first-of-its-kind analysis of \dspol, using a dataset of 8M posts from over 216K conversation threads collected over a 2.5-month period.
First, we perform a general characterization of \dspol, focusing on posting behavior and on how 4chan's unique features influence the way discussions proceed.
Next, we explore the types of content shared on \dspol, including third-party links and images, the use of hate speech,
and differences in discussion topics at the country level.
Finally, we show that \dspol's hate-filled vitriol is not contained within \dspol, or even 4chan, by measuring its effects on conversations taking place on other platforms, such as YouTube, via a phenomenon called ``raids.''

\descr{Contributions.} In summary, this paper makes several contributions.
First, we provide a large scale analysis of \dspol's posting behavior, showing the impact of 4chan's unique features, that \dspol users are spread around the world, and that, although posters remain anonymous, \dspol is filled with many different voices.
Next, we show that \dspol users post many links to YouTube videos, tend to favor ``right-wing'' news sources, and post a large amount of unique images.
Finally, we provide evidence that there are numerous instances of individual YouTube videos being ``raided,'' and provide a first metric for measuring such activity.

\descr{Paper Organization.} The rest of the paper is organized as follows. Next section provides an overview of 4chan and its main characteristics, then, Section~\ref{sec:related} reviews related work, while Section~\ref{sec:dataset} discusses our dataset. Then, Section~\ref{sec:characterization} and Section~\ref{sec:content-analysis} present, respectively, a general characterization and a content analysis of  \dspol. Finally, we analyze raids toward other services in Section~\ref{sec:raids}, while the paper concludes in Section~\ref{sec:conclusion}.

\section{4chan}\label{sec:background}

\url{4chan.org} is an imageboard site. A user, the ``original poster'' (OP), creates a new thread by posting a message, with an image attached, to a board with a particular topic. Other users can also post in the thread, with or without images,
and refer to previous posts by replying to or quoting portions of it.

\begin{figure}[t]
    \centering
    \includegraphics[width=0.985\columnwidth]{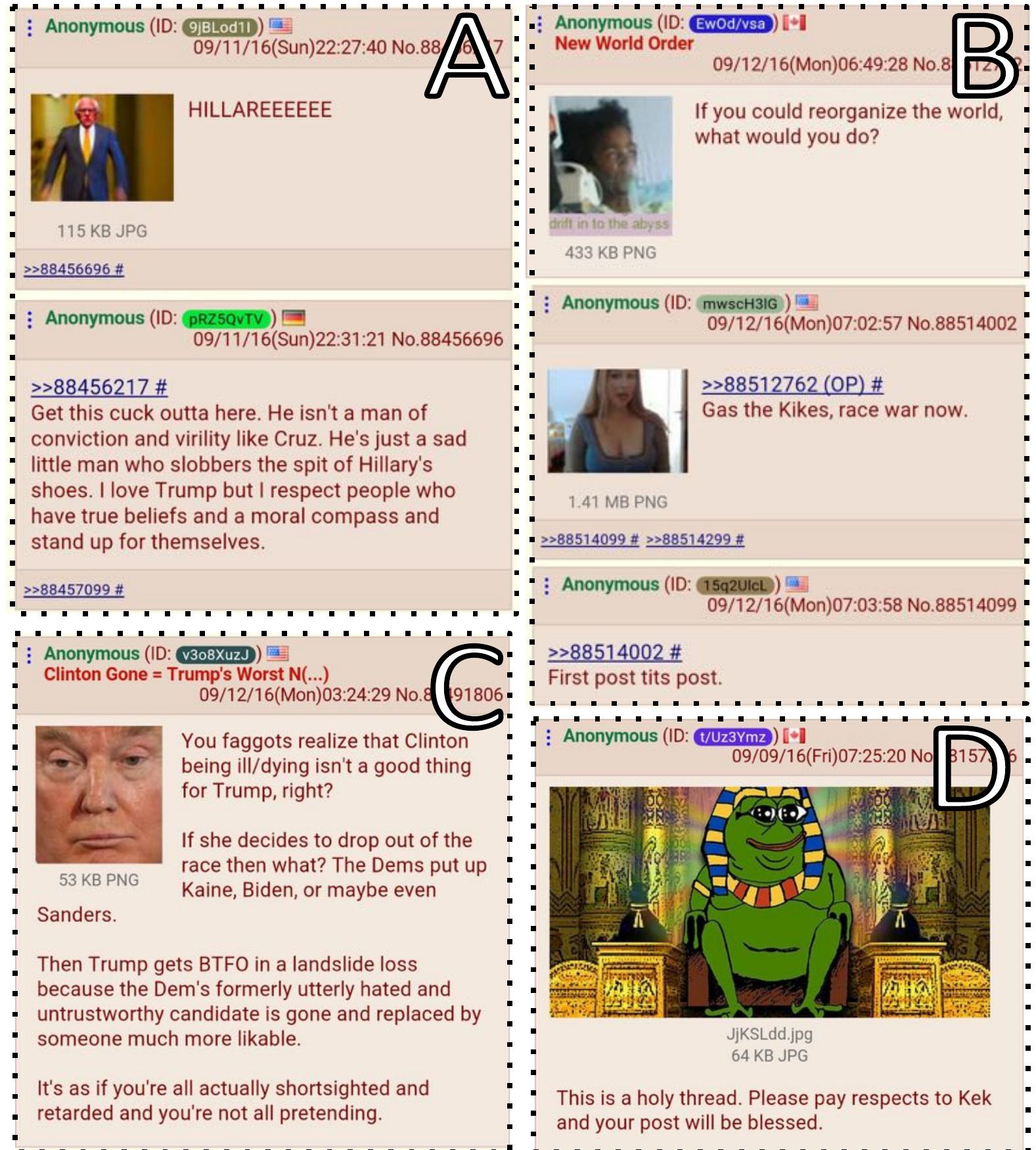}
    \vspace{-0.2cm}
    \caption{Examples of typical \dspolCCC threads. (A) illustrates the derogatory use of ``cuck'' in response to a Bernie Sanders image; (B) a casual call for genocide with an image of a woman's cleavage and a ``humorous'' response; (C) \dspolCCC's fears that a withdrawal of Hillary Clinton would guarantee Trump's loss; (D) shows {\em Kek}, the ``God'' of memes, via which \dspolCCC ``believes'' they influence reality.}
    \label{fig:pol-example-threads}
\end{figure}

\descr{Boards.}
As of January 2017, 4chan features 69 boards, split into 7 high level categories, e.g., Japanese Culture (9 boards) or Adult (13 boards).
In this paper, we focus on \dspol, the ``Politically Incorrect'' board.\footnote{\url{http://boards.4chan.org/pol/}}
Figure~\ref{fig:pol-example-threads} shows four typical \dspol threads.
Besides the content, the figure also illustrates the {\em reply} feature  (`>>12345' is a reply to post `12345'), as well as other concepts discussed below.
Aiming to create a baseline to compare \dspol to, we also collect posts from two other boards: ``Sports'' (\dssp) and ``International'' (\dsint).
The former focuses on sports and athletics, the latter on cultures, languages, etc.
We choose these two since they are considered ``safe-for-work'' boards,
and are, according to 4chan rules, more heavily moderated,
but also because they display the country flag of the OP, which we discuss next.

\descr{Anonymity.} Users do not need an account to read/write posts.
Anonymity is the default (and preferred) behavior, but
users can enter a name along with their posts, even though they can change it with each post if they wish. Naturally, anonymity here is meant to be with respect to other users,
not the site or the authorities, unless using Tor or similar tools.\footnote{In fact, moot (4chan's creator) reported turning server logs and other records over to the FBI. See~\url{http://www.thesmokinggun.com/buster/fbi/turns-out-4chan-not-lawless-it-seems}.}

\emph{Tripcodes} (hashes of user-supplied passwords) can be used to ``link'' threads from the same user across time,
providing a way to verify pseudo-identity.
On some boards, intra-thread trolling led to the introduction of \emph{poster IDs}.
Within a thread (and \emph{only} that thread), each poster is given a unique ID that appears along with their post, using a combination of cookies and IP tracking. This preserves anonymity, but mitigates low-effort sock puppeteering.
To the best of our knowledge, \dspol is currently the only board with poster IDs enabled.

\descr{Flags.} \dspol, \dssp, and \dsint also include, along with each post, the flag of the country the user posted from, based on IP geo-location.
This is meant to reduce the ability to ``troll'' users by, e.g., claiming to be from a country where an event is
happening (even though geo-location can obviously be manipulated using VPNs and proxies).

\descr{Ephemerality.}
Each board has a finite \emph{catalog} of threads. Threads are pruned after a relatively short period of time via a ``bumping system.''
Threads with the most recent post appear first, and
creating a new thread results in the one with the least recent post getting removed.
A post in a thread keeps it alive by bumping it up, however, to prevent a thread from never getting purged, 4chan implements {\em bump} and {\em image limits}. After a thread is bumped $N$ times or has $M$ images posted to it (with $N$ and $M$ being board-dependent),
new posts will no longer bump it up.
Originally, when a thread fell out of the catalog, it was permanently gone, however, an archive system for a subset of boards has  recently been implemented: once a thread is purged, its final state is archived for a relatively short period of time -- currently seven days.

\descr{Moderation.} 4chan's moderation policy is generally lax, especially on
\dspol.
So-called janitors, volunteers
periodically recruited from the user base,
can prune posts and threads, as well as recommend users to be banned by more ``senior'' 4chan employees.
Generally speaking, although janitors are not well respected by 4chan users and are often mocked for their perceived love for power,
they do contribute to 4chan's continuing operation,
by volunteering work on a site that is somewhat struggling to stay solvent~\cite{solvent}.

\section{Related Work}\label{sec:related}
While 4chan constantly attracts considerable interest in the popular press~\cite{wired,fortune},
there is very little scientific work analyzing its ecosystem.
To the best of our knowledge, the only measurement of 4chan is the work by~\cite{bernstein20114chan}, who study the ``random''
board on 4chan (\dsb), the original and most active board.
Using a dataset of 5.5M posts from almost 500K threads collected over a two-week period,
they focus on analyzing the anonymity and ephemerality characteristics of 4chan.
They find  that over 90\% of posts are made by anonymous
users, and, similar to our findings, that  the ``bump'' system affects threads' evolution,
as the median lifetime of a \dsb thread
is only 3.9mins (and 9.1mins on average).
Our work differs from~\cite{bernstein20114chan} in several aspects.
First, their study is focused on one board (\dsb) in a self-contained fashion, while we also measure how \dspol affects the rest of the Web (e.g., via raids). Second, their content analysis is primarily limited to a typology of thread types.
Via manual labeling of a small sample, they determined that 7\% of posts on \dsb are a ``call for action,'' which includes raiding behavior.
In contrast, our analysis goes deeper, looking at post contents and raiding in a quantitative manner.
Finally, using some of the features unique to \dspol, \dsint, and \dssp, we are also able to get a glimpse of 4chan's user demographics, which is only speculated about in~\cite{bernstein20114chan}.

\cite{2015arXiv151000240P} analyze the influence of anonymity
on aggression and obscene lexicon by comparing a few anonymous forums and social networks.
They focus on Russian-language platforms, and also include 2M words from 4chan, finding no correlation between anonymity and aggression.
In follow-up work~\cite{PotapovaG16},
4chan posts are also used to evaluate automatic verbal aggression detection tools.

Other researchers have also analyzed social media platforms, besides 4chan, characterized by
(semi-)anonymity and/or ephemerality.
\cite{correa2015} study the differences between content posted on anonymous and non-anonymous social media,
showing that linguistic differences between Whisper posts (anonymous)
and Twitter (non-anonymous) are significant, and they train
classifiers to discriminate them (with 73\% accuracy).
\cite{peddinti2014cloak} analyze users' anonymity choices during their activity on Quora, identifying categories of questions for which users are more likely to seek anonymity. They also
perform an analysis of Twitter to study the prevalence and behavior of so-called ``anonymous'' and ``identifiable'' users, as classified by Amazon Mechanical Turk workers, and find a correlation between content sensitivity and a user's choice to be anonymous.
\cite{HosseinmardiGHLM14} analyze user behavior on Ask.fm by building an ``interaction graph'' between 30K
profiles. They characterize users in
terms of positive/negative behavior and
in-degree/out-degree, and analyze the relationships between
these factors.

Another line of work focuses on detecting hate speech.~\cite{djuric2015hate} propose a word embedding based detection tool for hate speech on
Yahoo Finance.~\cite{www2016yahoopaper} also perform hate speech detection
on Yahoo Finance and News data, using a supervised classification methodology.
\cite{cheng2015trolls} characterize anti-social behavior in comments sections of a few popular websites and predict accounts on those sites that will exhibit anti-social behavior.
Although we observe some similar behavior from \dspol users, our  work is focused more on understanding the platform and organization of semi-organized campaigns of anti-social behavior, rather than identifying particular users exhibiting such behavior.

\section{Datasets}\label{sec:dataset}

On June 30, 2016, we started crawling 4chan using its JSON API.\footnote{\url{https://github.com/4chan/4chan-API}}
We retrieve \dspol's thread catalog every 5 minutes and
compare the threads that are currently live to those in the previously obtained catalog.
For each thread that has been purged, we retrieve a full copy from 4chan's archive,
which allows us to obtain the full/final contents of a thread.
For each post in a thread, the API returns, among other things, the post's number, its author (e.g., ``Anonymous''), timestamp, and contents of the post (escaped HTML).
Although our crawler does not save images, the  API also includes image metadata, e.g., the name the image is uploaded with, dimensions (width and height), file size, and an MD5 hash of the image.
On August 6, 2016 we also started crawling \dssp, 4chan's sports board, and on August 10, 2016 \dsint, the international board.
Table~\ref{tbl:dataset-overview} provides a high level overview of our datasets.
We note that for about 6\% of the threads, the crawler gets a 404 error:
from a manual inspection, it seems that this is due to ``janitors'' (i.e., volunteer moderators) removing threads for violating rules.

\begin{table}[t]
    \centering
\resizebox{0.95\columnwidth}{!}{%
    \begin{tabular}{ l r r r r }
        \hline
        & \multicolumn{1}{c}{\dspol} & \multicolumn{1}{c}{\dssp} & \multicolumn{1}{c}{\dsint} & \multicolumn{1}{c}{Total} \\
        \hline
        {\bf Threads} & 216,783 & 14,402 & 24,873 & 256,058 \\
        {\bf Posts} & 8,284,823 & 1,189,736 & 1,418,566 & 10,893,125 \\
        \hline
    \end{tabular}
    }
 \vspace{-0.15cm}
    \caption{Number of threads and posts crawled for each board.}
    \label{tbl:dataset-overview}
 \vspace{-0.15cm}

\end{table}

The analysis presented
in this paper considers data crawled until September 12,
2016, \emph{except} for the raids analysis presented later on, where we considered threads and YouTube comments
up to Sept. 25.
We also use a set of 60,040,275 tweets from Sept. 18 to Oct. 5, 2016 for a brief comparison in hate speech usage.
We note that our datasets are available to other researchers upon request.

\descr{Ethical considerations.}
Our study has obtained approval by the designated ethics officer at UCL.
We note that 4chan posts are typically anonymous, however, analysis of the activity generated by links on 4chan to other services could be potentially used to de-anonymize users.
To this end, we followed standard ethical guidelines~\cite{rivers2014ethical}, 
encrypting data at rest, and making no attempt to de-anonymize users.
We are also aware that content posted on \dspol is often highly offensive,
however, we do not censor content in order to provide a comprehensive analysis of \dspol,
but warn readers that the rest of this paper features language likely to be upsetting.

\section{General Characterization}\label{sec:characterization}

\subsection{Posting Activity in \dspolTTT}

\begin{figure}[t]
    \centering
    \includegraphics[width=0.85\columnwidth]{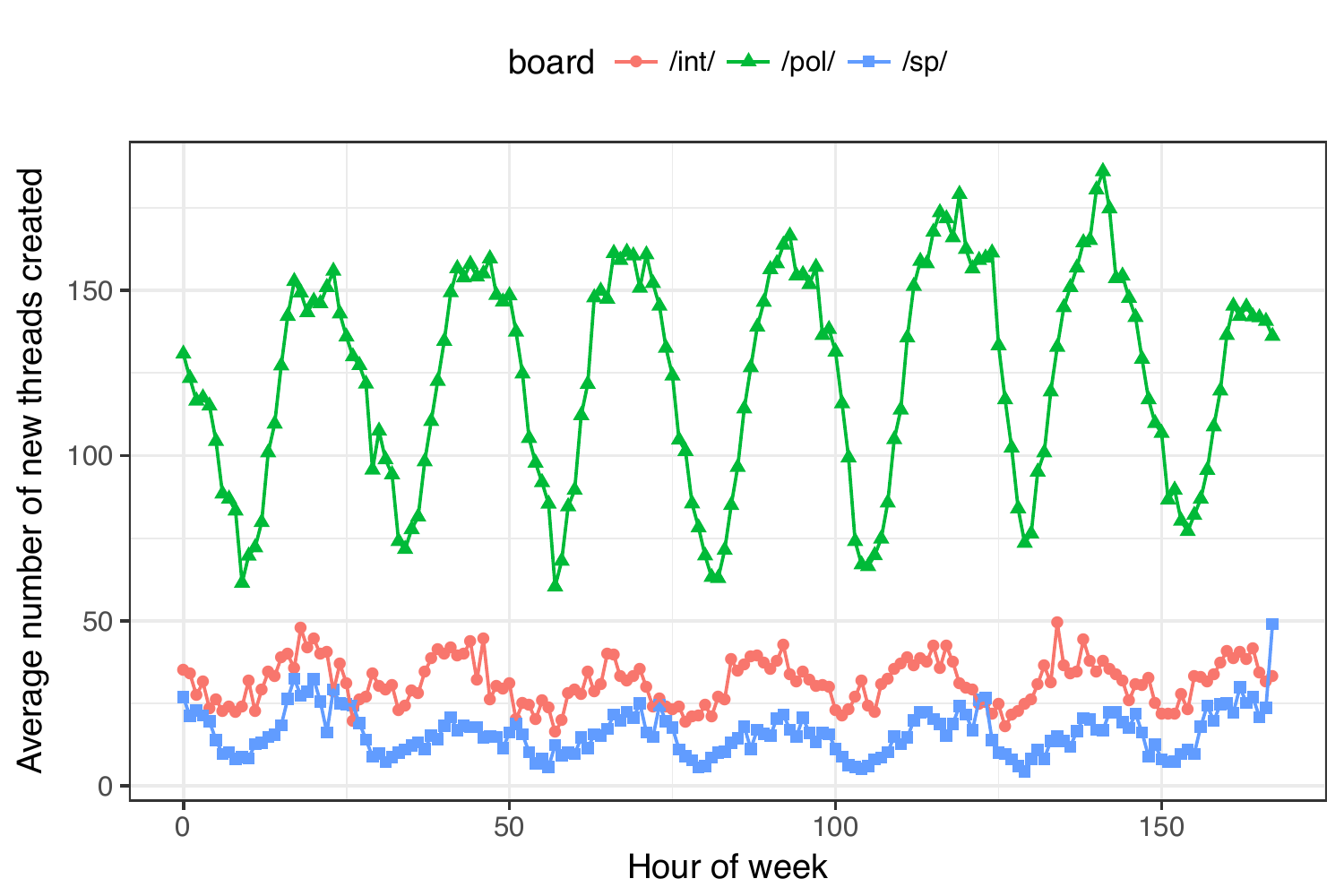}
    \vspace{-0.2cm}
    \caption{Average number of new threads per hour of the week.}
    \label{fig:mean-new-threads}
    \vspace{-0.05cm}
\end{figure}

\begin{figure}[t]
    \centering
    \includegraphics[width=0.99\columnwidth]{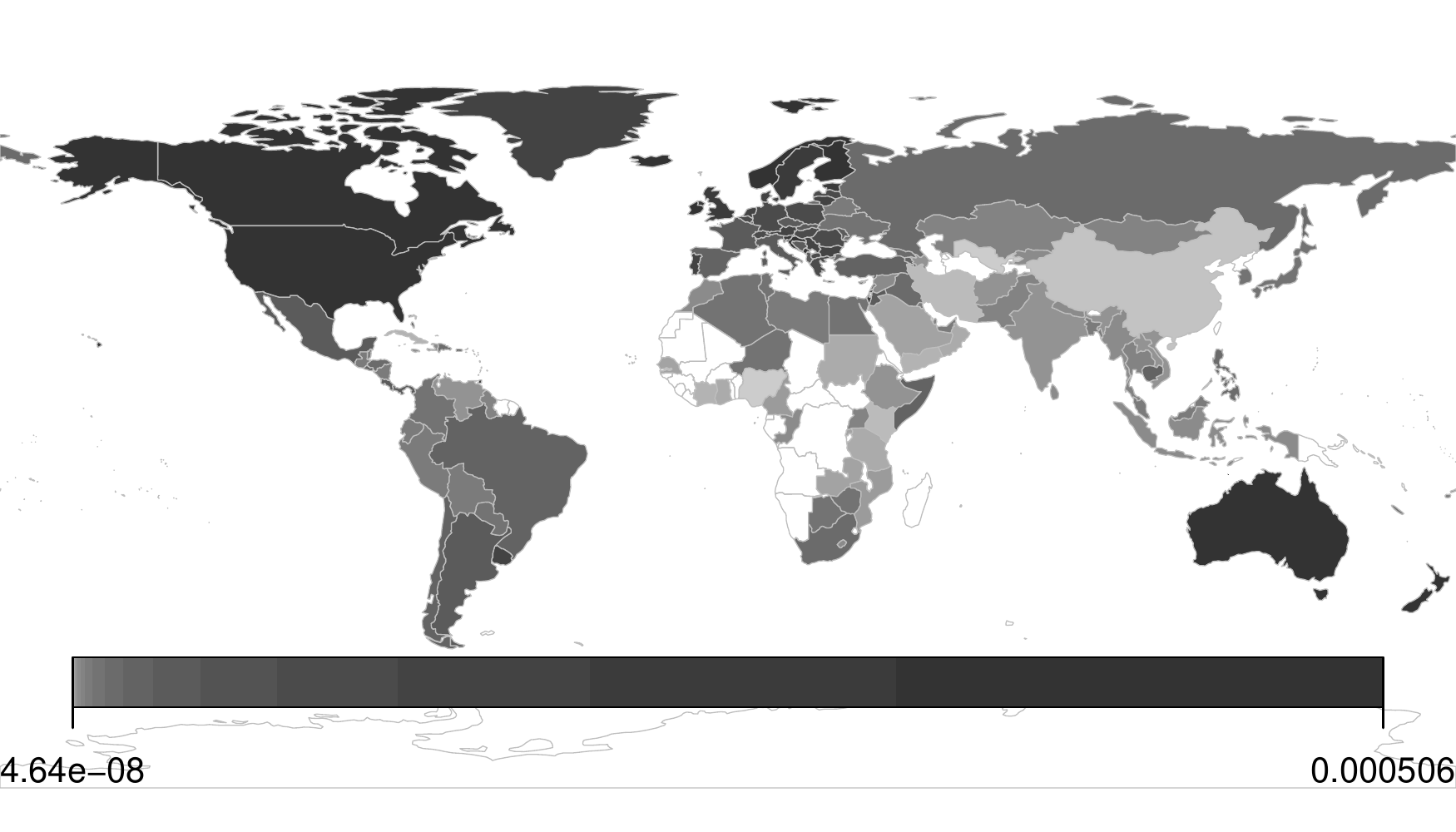}
    \vspace{-0.2cm}
    \caption{Heat map of the number of new \dspolCCC threads created per country, normalized by Internet-using population. The darker the country, the more participation in \dspolCCC it has, relative to its real-world Internet using population.}
    \label{fig:new-threads-by-country-normed}
    \vspace{-0.05cm}
\end{figure}

Our first step is a high-level examination of posting activity.
In Figure~\ref{fig:mean-new-threads}, we plot the average number of new threads created per hour of the week, showing that \dspol users create one order of magnitude more threads than \dsint and \dssp users at nearly all hours of the day. Then, Figure~\ref{fig:new-threads-by-country-normed} reports the number of new threads created per country,
normalized by the country's Internet-using population.\footnote{Obtained from \url{http://www.internetlivestats.com/internet-users/}}
Although the US dominates in total thread creation (visible by the timing of the diurnal patterns from Figure~\ref{fig:mean-new-threads}), the top 5 countries in terms of threads per capita are New Zealand, Canada, Ireland, Finland, and Australia.
4chan is primarily an English speaking board, and indeed nearly every post on \dspol is in English, but we still find that many non-English speaking countries -- e.g., France, Germany, Spain, Portugal, and several Eastern European countries -- are represented.
This suggests that although \dspol is considered an ``ideological backwater,'' it is surprisingly diverse in terms of international participation.

\begin{figure}[t]
    \centering
    \includegraphics[width=\columnwidth]{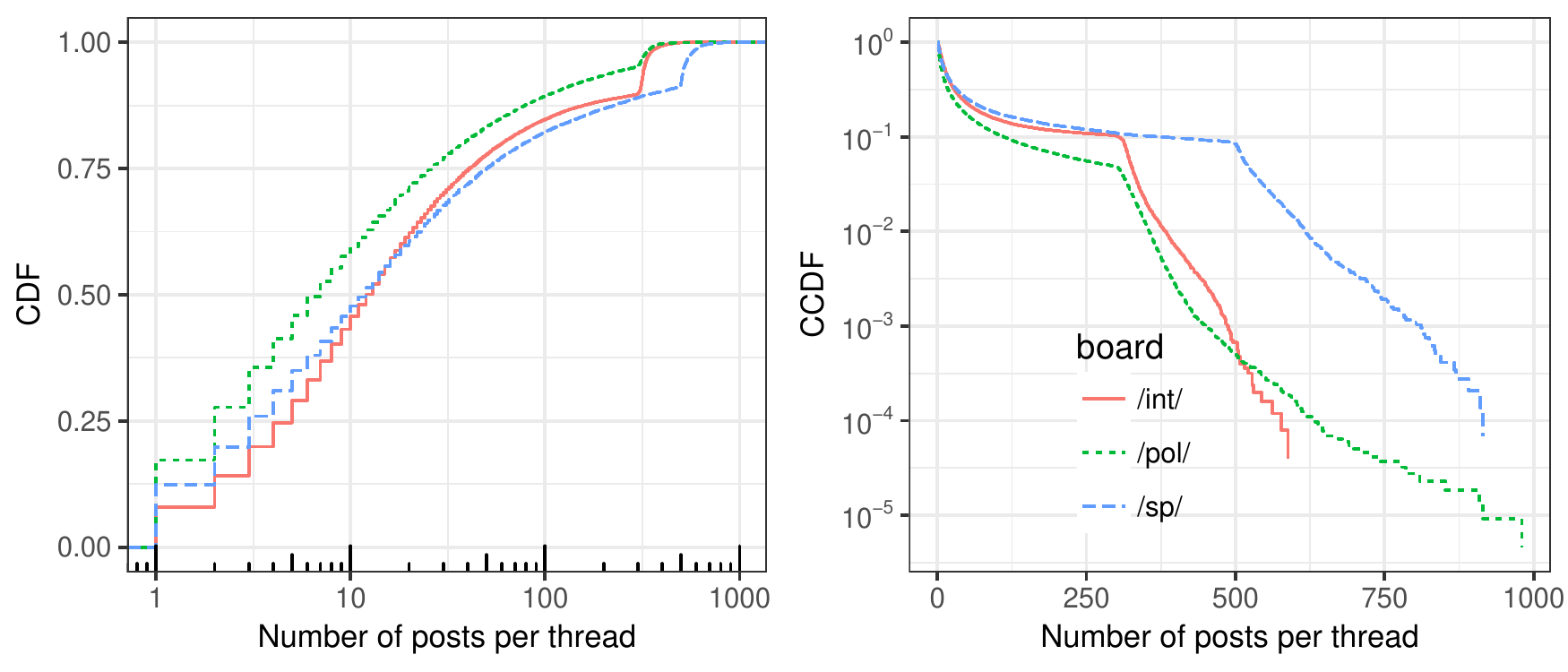}
    \vspace{-0.5cm}
    \caption{Distributions of the number of posts per thread on \dspolCCC, \dsintCCC, and \dsspCCC. We plot both the CDF and CCDF to show both typical threads as well as threads that reach the bump limit. Note that the bump limit for \dspolCCC and \dsintCCC is 300 at the time of this writing, while for \dsspCCC it is 500.}
    \label{fig:num-posts-per-thread}
    \vspace{-0.1cm}
\end{figure}

Next, in Figure~\ref{fig:num-posts-per-thread}, we plot the distribution of the number of posts per thread on \dspol, \dsint, and \dssp, reporting both the cumulative distribution function (CDF) and the complementary CDF (CCDF).
All three boards are skewed to the right, exhibiting quite different means (38.4, 57.1, and 82.9 for \dspol, \dsint, and \dssp, respectively) and medians (7.0, 12.0, 12.0) -- i.e., there are a few threads with a substantially higher number of posts.
One likely explanation for the average length of \dssp threads being larger is that users on \dssp make ``game threads'' where they discuss a professional sports game live, while it is being played.
The effects of the bump limit
are evident on all three boards.
The bump limit is designed to ensure that fresh content is always available, and Figure~\ref{fig:num-posts-per-thread} demonstrates this: extremely popular threads have their lives cut short earlier than the overall distribution would imply and are eventually purged.

\begin{figure*}[!htb]
    \centering
    \begin{minipage}{0.325\textwidth}
        \centering
        \includegraphics[width=\linewidth]{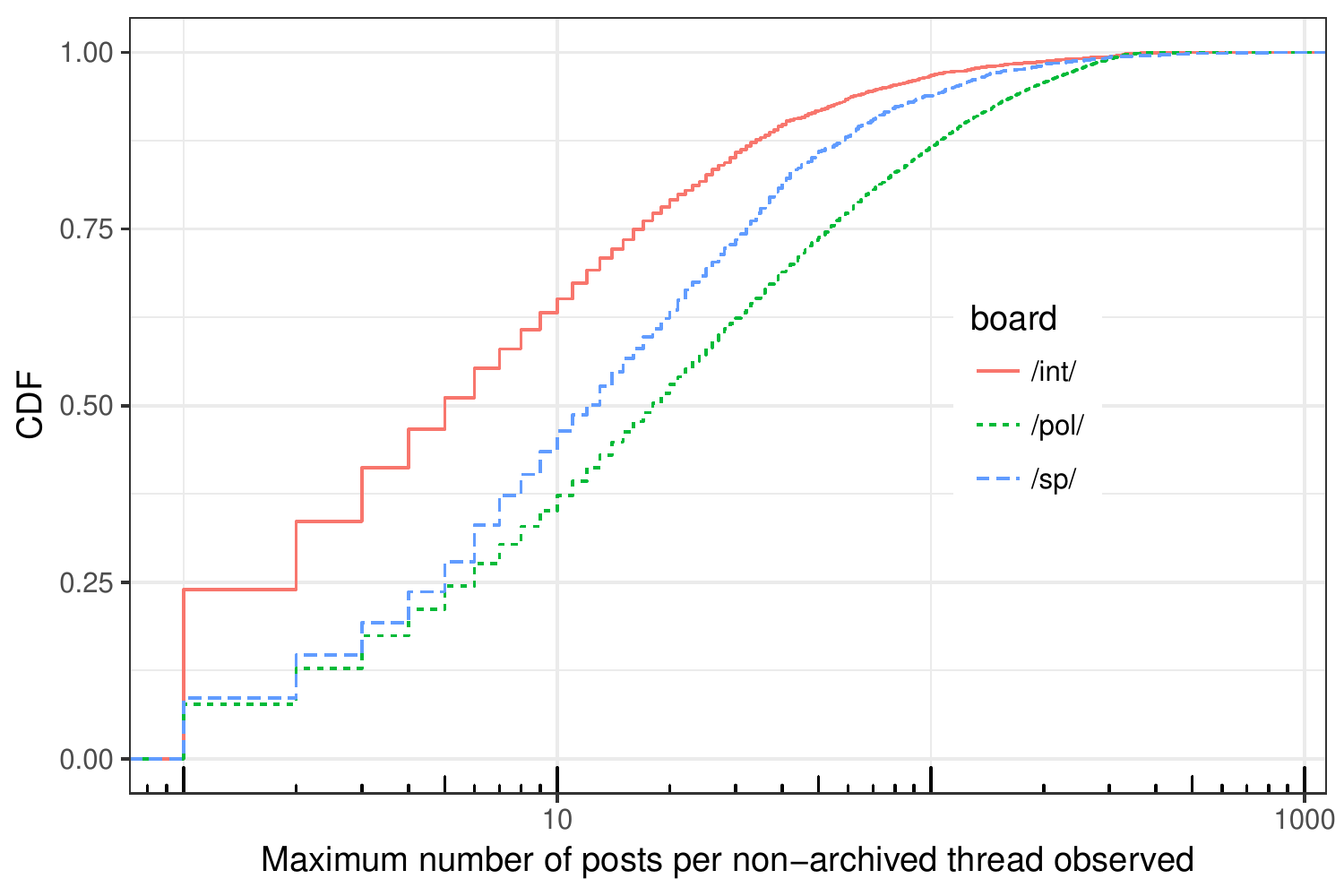}
        \vspace{-0.5cm}
        \caption{CDF of the number of posts for non-archived threads (i.e., likely deleted).}
        \label{fig:max-posts-per-non-archived-thread}
    \end{minipage}\hfill%
    \begin{minipage}{0.325\textwidth}
        \centering
        \includegraphics[width=\linewidth]{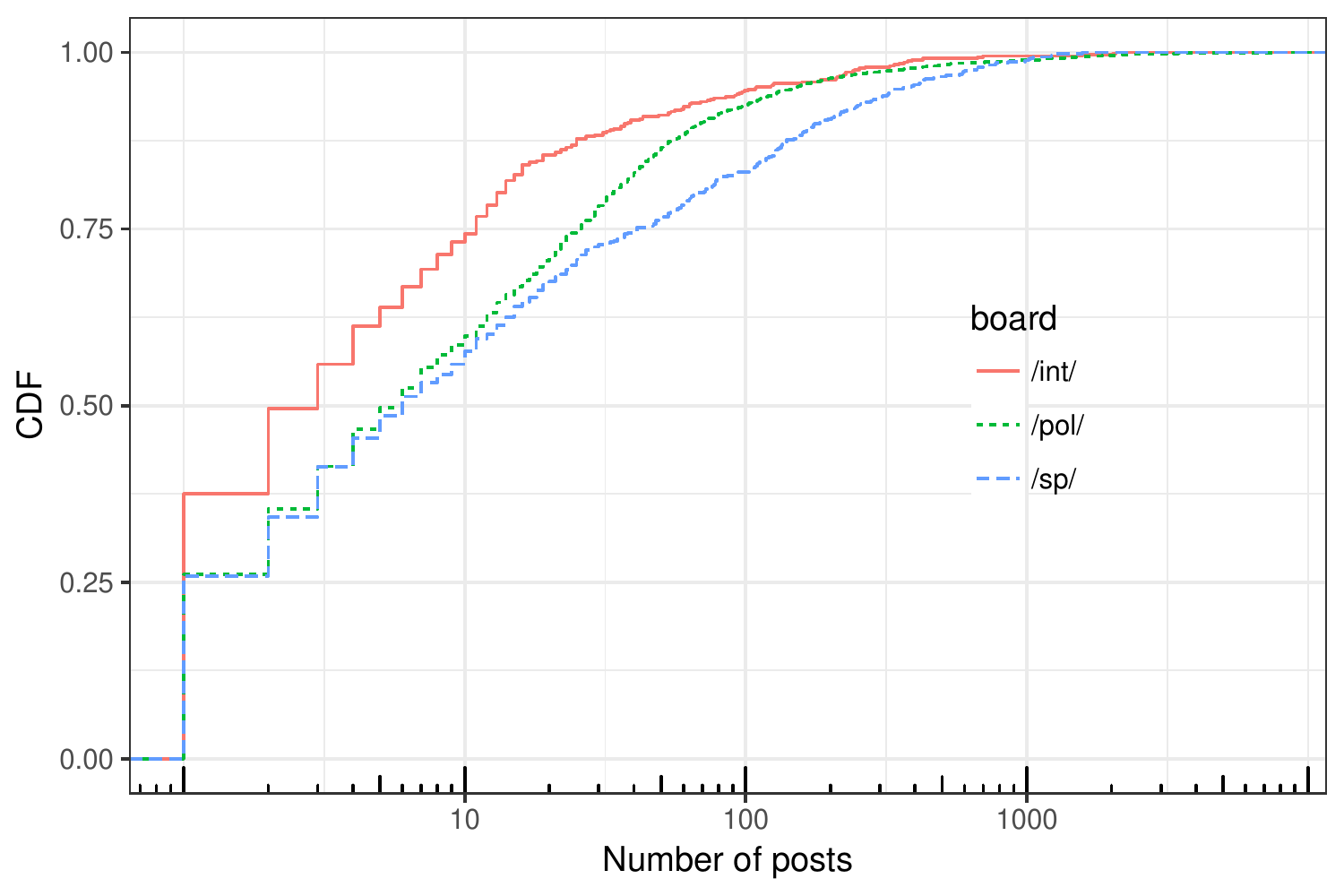}
        \vspace{-0.5cm}
        \caption{CDF of the number of posts per unique tripcode.}
        \label{fig:posts-per-tripcode}
    \end{minipage}\hfill%
    \begin{minipage}{0.325\textwidth}
        \centering
        \includegraphics[width=\linewidth]{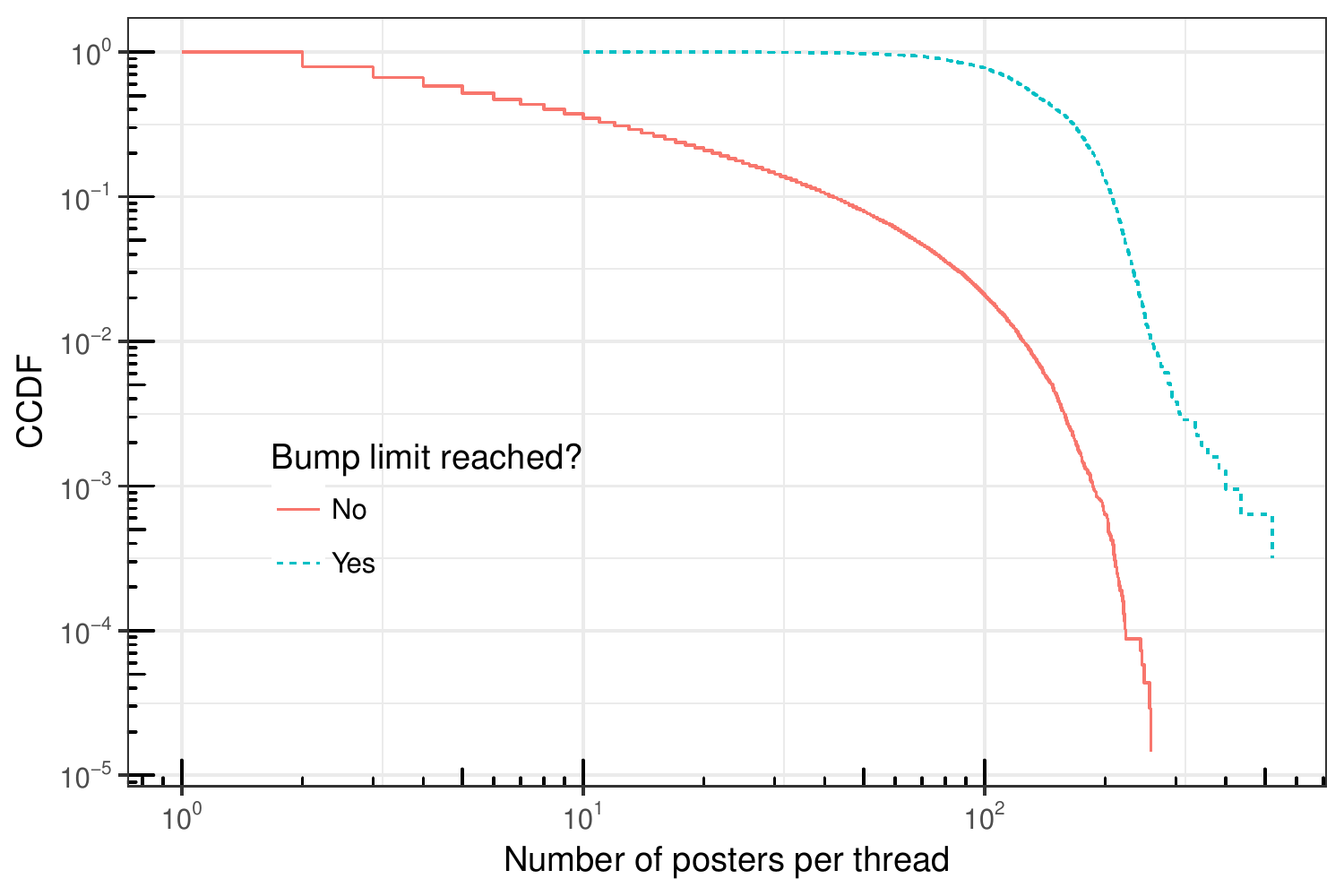}
        \vspace{-0.5cm}
        \caption{CCDF of the number of unique posters per thread.}
        \label{fig:unique-users-per-thread}
    \end{minipage}%
    \vspace{-0.05cm}
\end{figure*}

We then investigate how much content actually violates the rules of the board.
In Figure~\ref{fig:max-posts-per-non-archived-thread}, we plot the CDF of the maximum number of posts per thread observed via the \dspol catalog, but for which we later receive a 404 error when retrieving the archived version -- i.e., threads that have  been deleted by a janitor or moved to another board.
Surprisingly, there are many ``popular'' threads that are deleted, as the median number of posts in a deleted \dspol thread is around 20,
as opposed to 7 for the threads that are successfully archived.
For \dsint, the median number of posts in a deleted thread (5) is appreciably lower than in archived threads (12).
This difference is likely due to: 1)~\dsint moving much slower than \dspol, so there is enough time to delete threads before they become overly popular, and/or 2)~\dspol's relatively lax moderation policy, which allows borderline threads to generate many posts before they end up ``officially'' violating the rules of the board.

\subsection{Tripcodes, Poster IDs, and Replies}\label{sec:tripcodes}

Next, we aim to shed light on 4chan's user base. This task is not trivial, since, due to the site's anonymous and ephemeral
nature, it is hard to build a unified network of user interactions. However, we leverage 4chan's pseudo-identifying attributes --
i.e., the use of tripcodes and poster IDs -- to provide an overview of both micro-level interactions and individual poster behavior over time.

Overall, we find 188,849 posts with a tripcode attached across \dspol (128,839 posts), \dssp (42,431), and \dsint (17,578) -- out of the 10.89M total posts in our dataset (Table~\ref{tbl:dataset-overview}).
Note that unique tripcodes do not necessarily correspond to unique users, since users can use any number of tripcodes.
Figure~\ref{fig:posts-per-tripcode} plots the CDF of posts per unique tripcode, for each of the three boards, showing that the median and mean are 6.50 and 36.08, respectively.
We observe that 25\% of tripcodes (over 30\% on \dsint) are only used once,
and that, although \dspol has many more posts overall, \dssp has more active ``tripcode users'' -- about 17\% of tripcodes on \dssp are associated to at least 100 posts, compared to about 7\% on \dspol.

Arguably, the closest we can get to estimating how unique users are engaged in 4chan threads is via poster IDs.
Unfortunately, these are not available from the JSON API once a thread is archived, and we decided to use them only a few weeks into our data collection. However, since the HTML version of archived threads \emph{does} include poster IDs, we started collecting HTML on August 17, 2016,
obtaining it for the last 72,725 (33\%) threads in our dataset.

Figure~\ref{fig:unique-users-per-thread} plots the CCDF of the number of unique users per \dspol thread, broken up into threads that reached the bump limit and those that did not.
The median and mean number of unique posters in threads that reached the bump limit was 134.0 and 139.6, respectively.
For typical threads (those that did not reach the bump limit), the median and mean is much lower -- i.e., 5.0 and 14.76 unique posters per thread.
This shows that, even though 4chan is anonymous, the most popular threads have ``many voices.''
Also recall that, in 4chan, replying to a particular post entails users referencing another post number \texttt{N} by adding \texttt{>>N} in their post, and the standard UIs then treat it as a reply.
This is different from simply posting in a thread:
users are \emph{directly} replying to a specific post (not necessarily the post the OP started the thread with),
with the caveat that one can reply to the same post multiple times and to multiple posts at the same time.

 \begin{figure}[t]
 	\centering
 	\includegraphics[width=0.8\columnwidth]{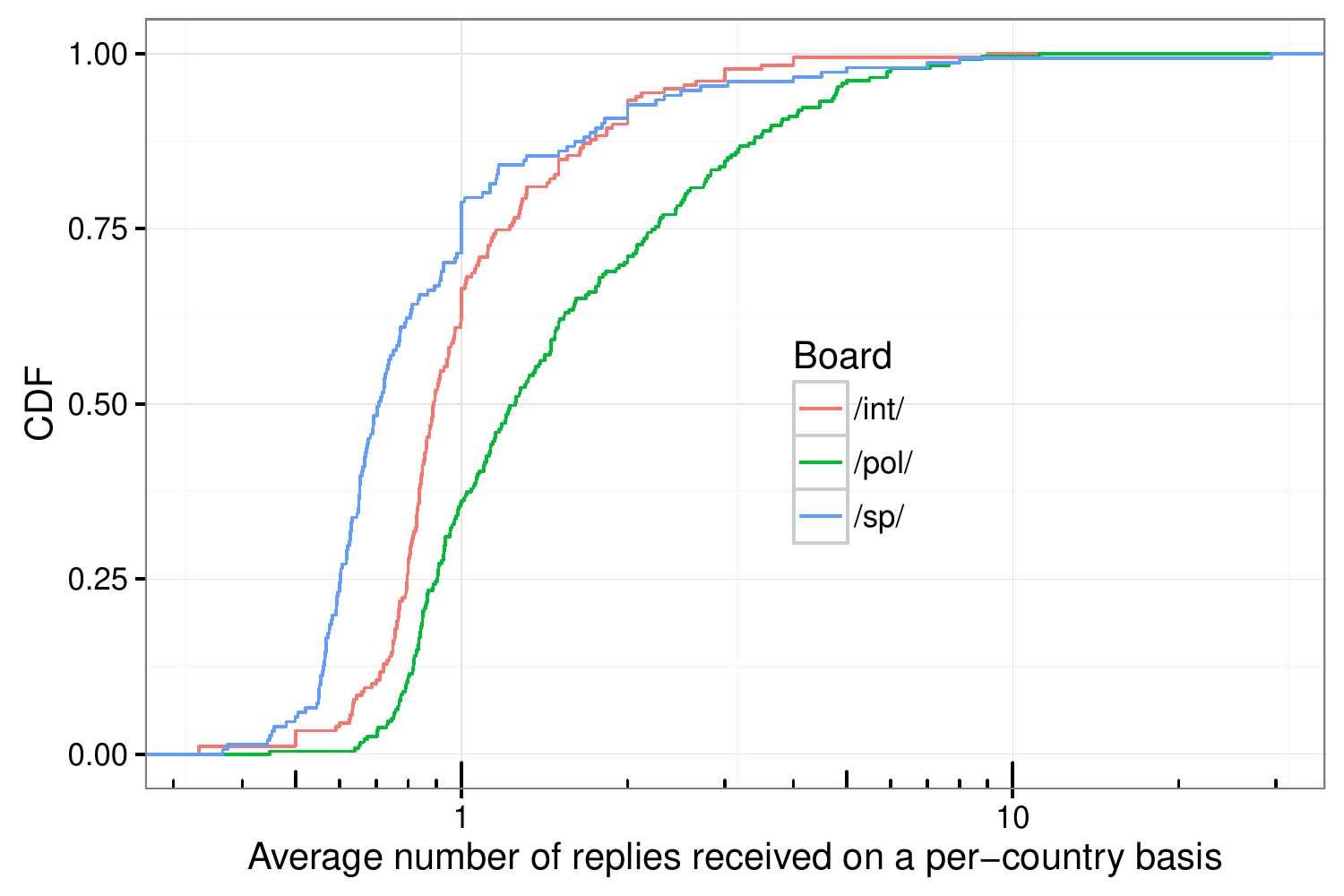}
 	\vspace{-0.2cm}
 	\caption{Distribution of the average number of replies received per country, per board.}
 	\label{fig:mean-replies-per-country}
 	\vspace{-0.2cm}
 \end{figure}

We look at this reply functionality in 4chan to assess how engaged users are with each other. First, we find that 50-60\% of posts never receive a direct reply across all three boards (\dsint: 49\%, \dspol: 57\%, \dssp: 60\%).
Taking the posts with no replies into account, we see that on average \dspol (0.83) and \dsint (0.80) have many more replies per post than \dssp ($0.64$), however, the standard deviation on \dspol is much higher (\dspol: 2.55, \dsint: 1.29, \dssp: 1.25).

Next, in Figure~\ref{fig:mean-replies-per-country}, we plot the CDF of the average number of replies per poster per board, aggregated by the country of the poster, i.e., the distribution of mean replies per country.
The top-10 countries (with at least 1,000 posts) per average number of replies
-- Table~\ref{tbl:top-country-replies} lets us zoom on the tail end of this ``replies-per-post'' per country distribution.
On average, while \dspol posts are likely to receive more replies than \dssp and \dsint posts, the distribution is heavily skewed towards certain countries.
Although deeper analysis of these differences is beyond the scope of this paper, we highlight that,
for some of the countries, the ``rare flag'' meme may be responsible for receiving more replies. I.e., users will respond to a post by an uncommonly seen flag. For other countries, e.g., Turkey or Israel, it might be the case that these are either of particular interest to \dspol, or are quite adept at trolling \dspol into replies (we note that our dataset covers the 2016 Turkish coup attempt and \dspol has a love/hate relationship with Israel).

Finally, we note that, unlike many other social media platforms, there is no other interaction system applied to posts on 4chan besides replies (e.g., no liking, upvoting, starring, etc.).
Thus, the only way for a user to receive validation from (or really any sort of direct interaction with) other users is to entice them to reply, which might encourage users to craft as inflammatory or controversial posts as possible.

\begin{table}[!t]
\centering 
  
\resizebox{0.99\columnwidth}{!}{%
\centering 

\begin{tabular}{cc|cc|cc} 
\hline
\multicolumn{2}{c|}{\dspol} & \multicolumn{2}{c|}{\dsint} & \multicolumn{2}{c}{\dssp}\\ 
Country & Avg. Replies & Country & Avg. Replies & Country & Avg. Replies \\
\hline
China & 1.57 & Thailand & 1.13 & Slovenia & 0.91 \\
Pakistan & 1.42 & Algeria & 1.12 & Japan & 0.84 \\
Japan & 1.35 & Jordan & 1.04 & Bulgaria & 0.81 \\
Egypt & 1.33 & S. Korea & 1.02 & Sweden & 0.75 \\
Tri. \& Tob. & 1.28 & Ukraine & 1.00 & Israel & 0.74 \\
Israel & 1.27 & Viet Nam & 0.97 & Argentina & 0.72 \\
S. Korea & 1.20 & Tunisia & 0.97 & India & 0.72 \\
Turkey & 1.18 & Israel & 0.97 & Greece & 0.72 \\
UAE & 1.20 & Hong Kong & 0.92 & Puerto Rico & 0.70 \\
Bangladesh & 1.15 & Macedonia & 0.91 & Australia & 0.68 \\

\hline
\end{tabular} 
}
\vspace{-0.2cm}
\caption{The top 10 countries (with at least 1,000 posts) in terms of direct replies received per post for each board in our dataset.} 
  \label{tbl:top-country-replies} 
\vspace{-0.2cm}
\end{table}

\section{Analyzing Content}\label{sec:content-analysis}
In this section, we present an exploratory analysis of the content posted on \dspol.
First, we analyze the types of media (links and images) shared on the board, then,
we study the use of hate words, and show how \dspol users can be clustered into meaningful geo-political regions
via the wording of their posts.

\subsection{Media Analysis}

\begin{figure}
    \centering
    \includegraphics[width=0.95\columnwidth]{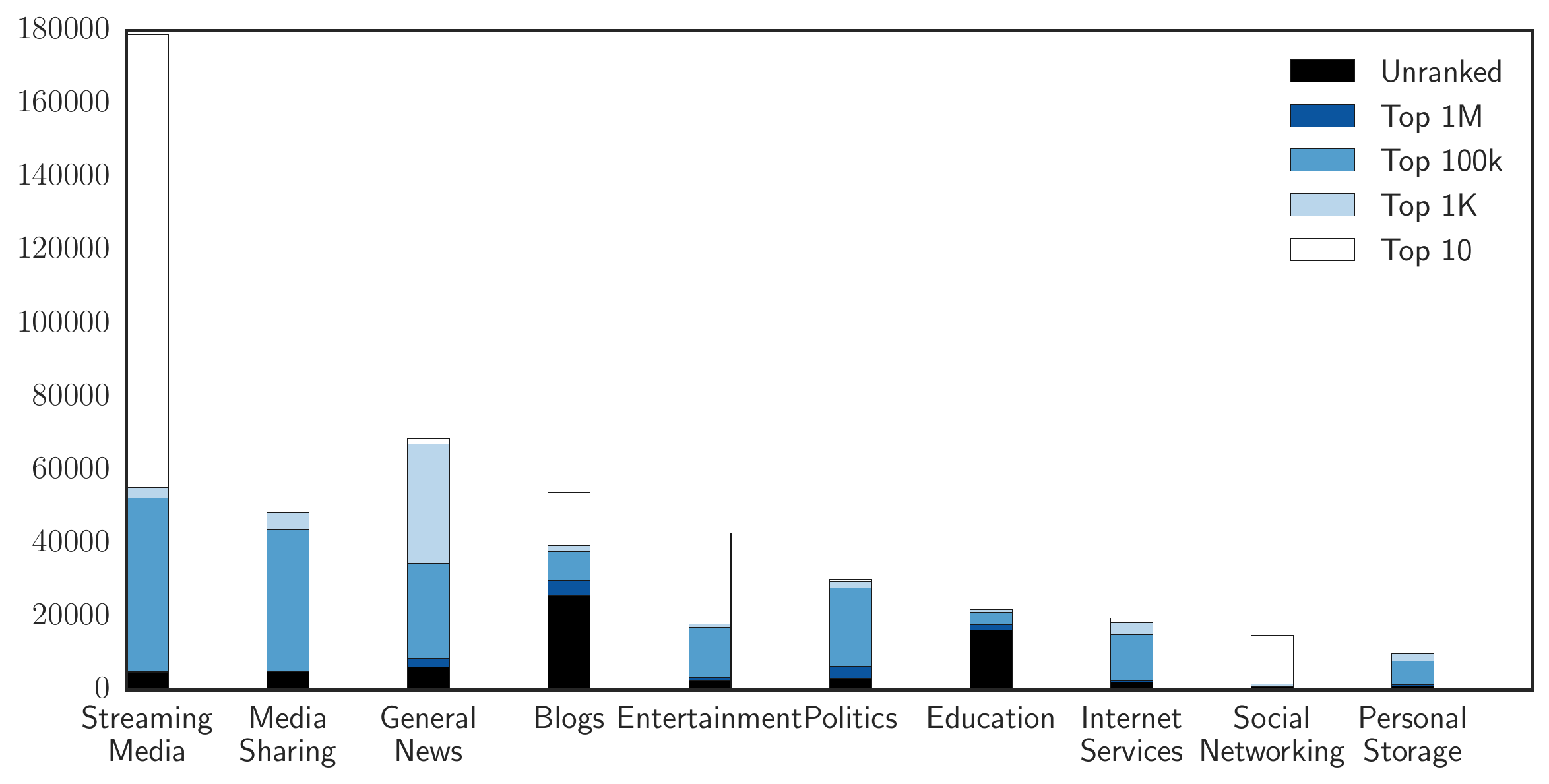}
\vspace{-0.1cm}
\caption{Distribution of different categories of URLs posted in \dspolCCC, together with the
Alexa ranking of their domain.}
    \label{fig:pol-urls}
    \vspace{-0.15cm}
\end{figure}

\descr{Links.} As expected, we find that \dspol users often post links to external content, e.g.,
to share and comment on news and events.
(As we discuss later, they also do so to identify
and coordinate targets for hate attacks on other platforms.)
To study the nature of the URLs posted on \dspol, we use
McAfee SiteAdvisor,\footnote{\url{https://www.siteadvisor.com/}}
which, given a URL, returns its category -- e.g., ``Entertainment'' or ``Social Networking.''
We also measure the popularity of the linked websites, using Alexa ranking.\footnote{\url{http://www.alexa.com/}}
Figure~\ref{fig:pol-urls} plots the distribution of categories of URLs posted in
\dspol, showing that ``Streaming Media'' and ``Media Sharing'' are
the most common, with YouTube playing a key role.
Interestingly, for some categories, URLs mostly belong to very popular domains, while others, e.g., ``General News,'' include a large number of less popular sites.

The website most linked to on \dspol is YouTube, with over an order of magnitude more URLs posted than the next two sites, Wikipedia and Twitter,
followed by Archive.is, a site that lets users take on-demand ``snapshots'' of a website,
which is often used on \dspol to record content -- e.g., tweets, blog posts, or news stories -- users feel might get deleted.
The 5th and 6th most popular domains are Wikileaks and pastebin,
followed by DonaldJTrump.com. Next, news sites start appearing, including the DailyMail and
Breitbart, which are right-wing leaning news outlets.
It is interesting to observe that some of the most popular news sites on a global level, e.g., CNN, BBC, and The Guardian,
appear well outside the top-10 most common domains.
On a board like \dspol,
which is meant to focus on politics and current events, this underlines
the polarization of opinions expressed by its users.

\begin{figure*}[!htb]
    \centering
    \begin{minipage}{0.315\textwidth}
        \centering
        \includegraphics[width=\linewidth]{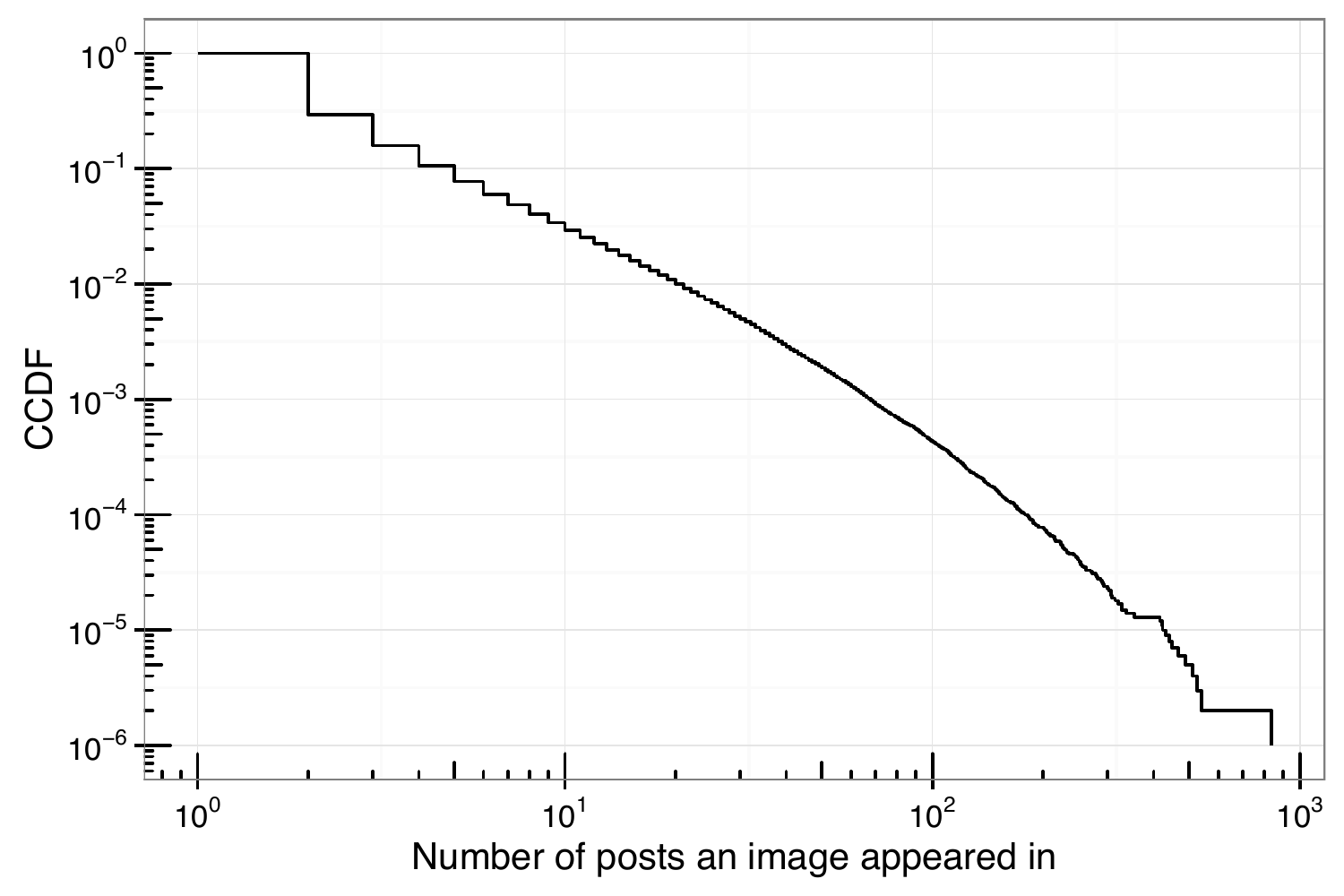}
        \vspace{-0.5cm}
        \caption{CCDF of the number of posts exact duplicate images appeared in on \dspolCCC.}
        \label{fig:image-reuse-ccdf}
    \end{minipage}\hfill%
    \begin{minipage}{0.315\textwidth}
        \centering
        \includegraphics[width=\linewidth]{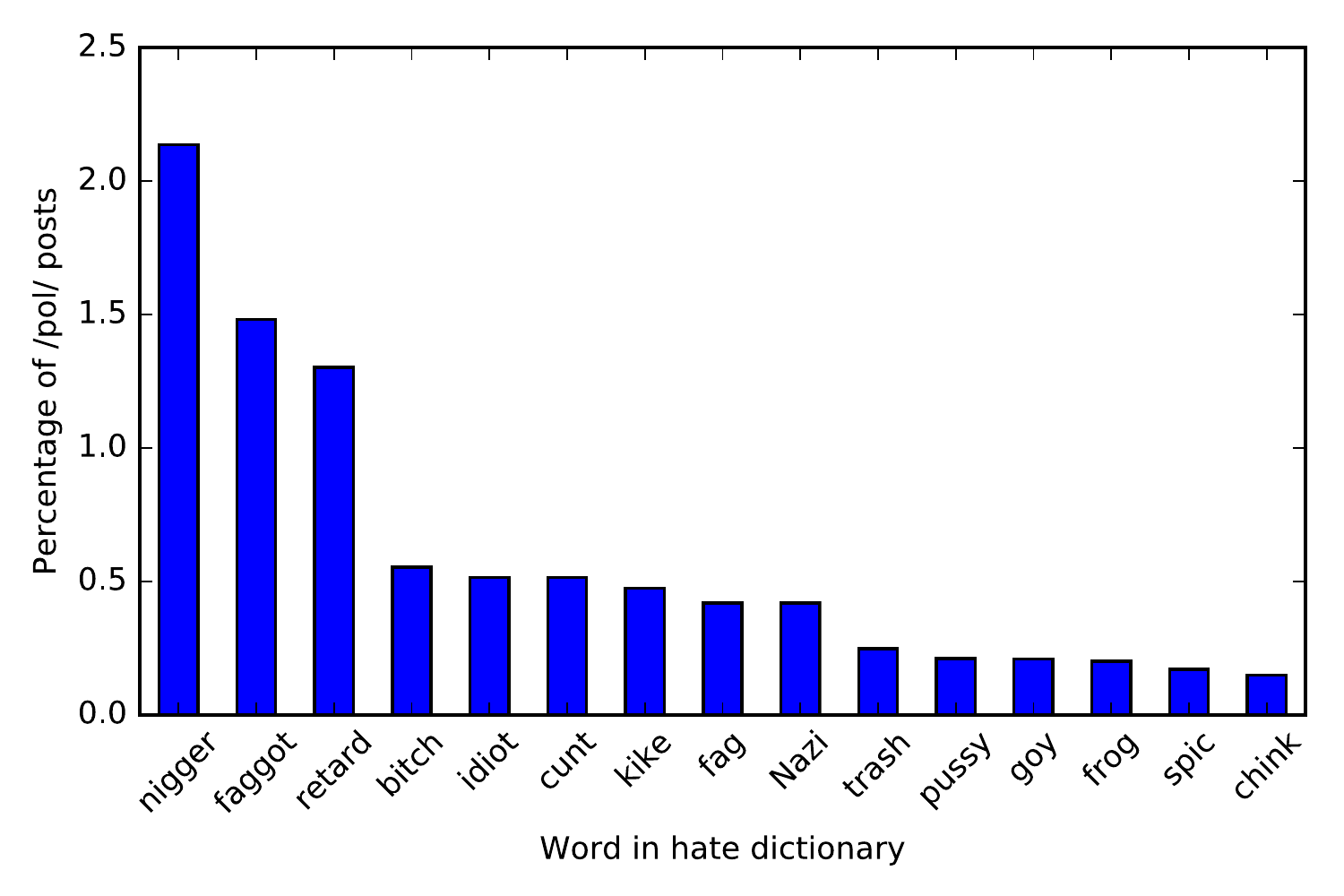}
        \vspace{-0.5cm}
        \caption{Percentage of posts on \dspolCCC the top 15 most popular hate words appear in.}
        \label{fig:popular-hate-words}
    \end{minipage}\hfill%
    \begin{minipage}{0.345\textwidth}
        \centering
        \vspace{0.15cm}
        \includegraphics[width=\linewidth]{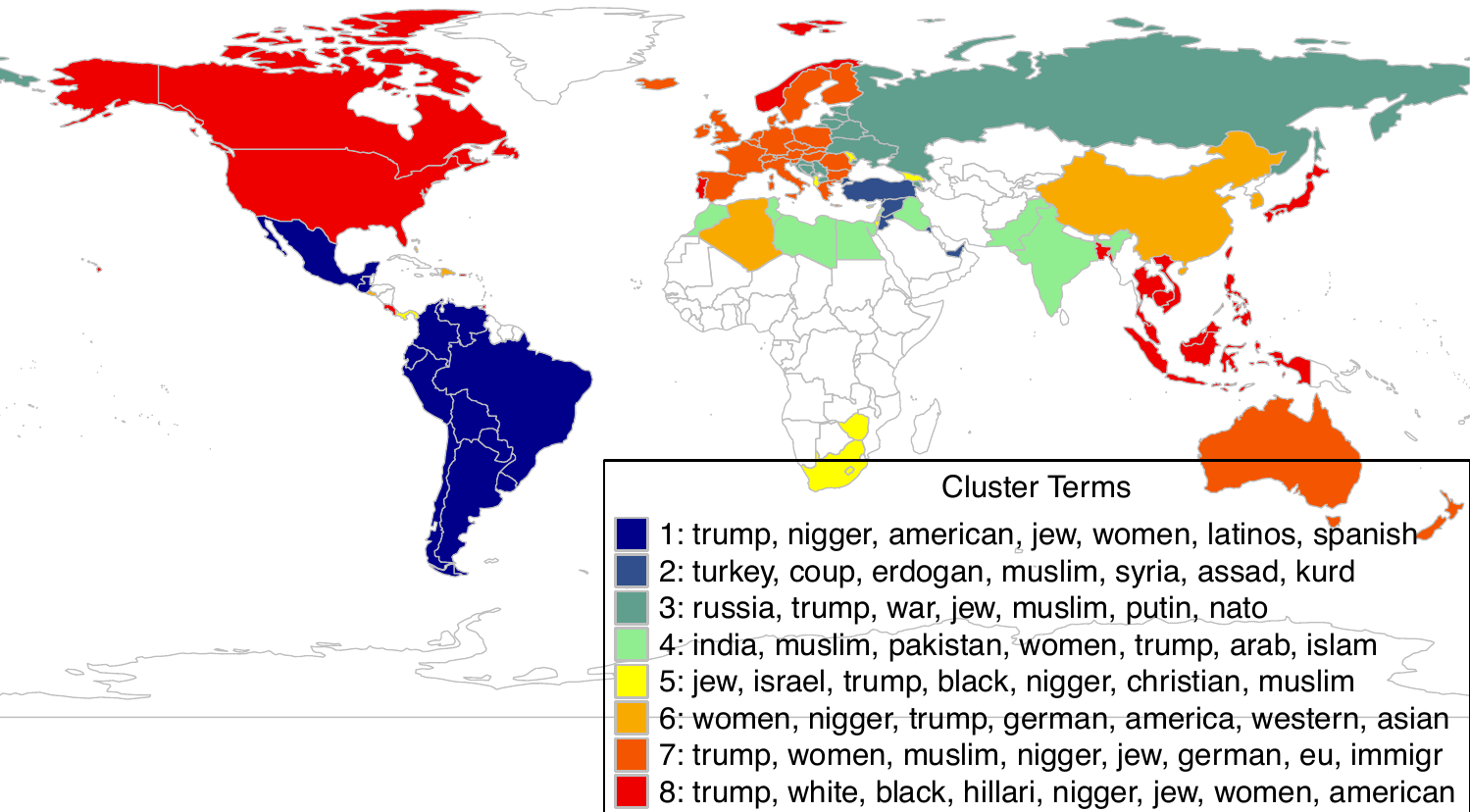}
        \vspace{-0.35cm}
        \caption{World map colored by content analysis based clustering.}
        \label{fig:country-cluster-map}
    \end{minipage}%
    \vspace{-0.05cm}
\end{figure*}

 \begin{figure}[t]
 	\centering
 	\includegraphics[width=0.45\columnwidth]{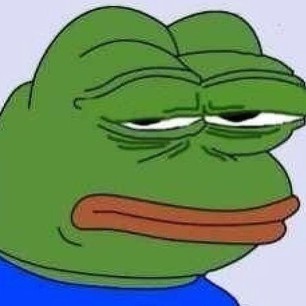}
 	\caption{The most popular image on \dspolCCC during our collection period, perhaps the least rare Pepe.}
 	\label{fig:most-popular-pepe}
 \end{figure}

\descr{Images.} 4chan was designed as an imageboard site, where users share images
along with a message. Therefore, although some content will naturally be ``reposted'' (in fact, memes
are almost by definition going to be posted numerous times~\cite{ferrara2013memes}), we expect  \dspol to generate large amounts of original content.
To this end, we count the number of unique images posted on \dspol during our observation period, finding 1,003,785 unique images (almost 800GB) out of a total 2,210,972 images (45\%).
We also plot the CCDF of the number of posts in which each unique image appears, using the image hash (obtained from the JSON API)
as a unique identifier, in Figure~\ref{fig:image-reuse-ccdf}.
Although the plot is only a \emph{lower} bound on image reuse (it only captures \emph{exact} reposts), we note that the majority (about 70\%) of images are only posted once, and nearly 95\% no more than 5 times.
That said, there is a very long tail, i.e., a few select images become what we might deem ``successful memes.''
This is line with 4chan's reputation for creating memes, and a meme is such only if it is seen many times.
Indeed, the most popular image on \dspol appears 838 times in our dataset, depicting what we might consider the least rare ``Pepe '' -- see Figure~\ref{fig:most-popular-pepe}.
Note that the \emph{Pepe the Frog} meme was recently declared a hate symbol by the Anti-Defamation League~\cite{adl-pepe}, but
of the 10 Pepe images appearing in the top 25 most popular images on \dspol, none seem to have an obvious link to hate.
While Figure~\ref{fig:most-popular-pepe} is clearly the most common of Pepes, we have included a collection of somewhat rarer Pepes in Appendix~\ref{sec:pepes}.

Even with a conservative estimation, we find that \dspol users posted over 1M unique images in 2.5 months, the majority of which were either original content or sourced from outside \dspol.
This seems to confirm that the constant production of new content may be one of the reasons \dspol is at the heart of the hate movement on the Internet~\cite{dailybeast2015pol}.

\subsection{Text Analysis}

 \descr{Hate speech.} \dspol is generally considered a ``hateful'' ecosystem, however, \emph{quantifying} hate is a non-trivial task. One possible approach is to perform sentiment analysis~\cite{pang2008opinion} over the posts in order to identify positive vs. negative attitude,
but this is difficult since the majority of \dspol posts (about 84\%) are either neutral or negative.
As a consequence, to identify hateful posts we use the \emph{hatebase} dictionary, a crowdsourced list of more than 1,000 terms from around the world that indicate hate when referring to a third person.\footnote{\url{https://www.hatebase.org}}
 We also use the NLTK framework\footnote{\url{http://www.nltk.org}} to identify these words in various forms (e.g., ``retard'' vs ``retarded'').
Our dictionary-based approach identifies posts that \emph{contain} hateful terms, but there might be cases where the context might not exactly be ``hateful'' (e.g., ironic usage). Moreover, \emph{hatebase} is a crowdsourced database, and is not perfect. To this end, we manually examine the list and remove a few of the words that are clearly ambiguous or extremely context-sensitive (e.g., ``india'' is a variant of ``indio,'' used in Mexico to refer to someone of Afro-Mexican origin, but is likely to be a false positive confused with the country India in our dataset). Nevertheless, given the nature of \dspol, the vast majority of posts likely use these terms in a hateful manner.

Despite these caveats, we can use this approach to provide an idea of how prevalent hate speech is on \dspol. We find that 12\% of \dspol posts contain hateful terms, which is substantially higher than in \dssp (6.3\%) and \dsint (7.3\%).  In comparison, analyzing our sample of tweets reveals just how substantially different \dspol is from other social media: only 2.2\% contained a hate word. In Figure~\ref{fig:popular-hate-words}, we also report the percentage of \dspol posts in which the top 15 most ``popular'' hate words from the hatebase dictionary appear.
``Nigger'' is the most popular hate word, used  in  more than 2\% of posts, while ``faggot'' and ``retard'' appear in over 1\% of posts.  To get an idea of the \emph{magnitude} of hate, consider that ``nigger'' appears in 265K posts, i.e., about 120 posts an hour.
After the top 3 hate words, there is a sharp drop in usage, although we see a variety of slurs.
These include ``goy,'' which is a derogatory word used by Jewish people to refer to non-Jewish people.
In our experience, however, we note that ``goy'' is used in an \emph{inverted} fashion on \dspol, i.e., posters call other posters ``goys'' to imply that they are submitting to Jewish ``manipulation'' and ``trickery.''

\descr{Country Analysis.}
Next, we explored how hate speech differs by country.
We observe clear differences in the use of hate speech, ranging from around 4.15\% (e.g., in Indonesia, Arab countries, etc.) to around 30\% of posts (e.g., China, Bahamas, Cyprus), while the majority of the 239 countries in our dataset feature hate speech in 8\%--12\% of their posts.

\begin{figure}[t]
 	\centering
 	\includegraphics[width=\columnwidth]{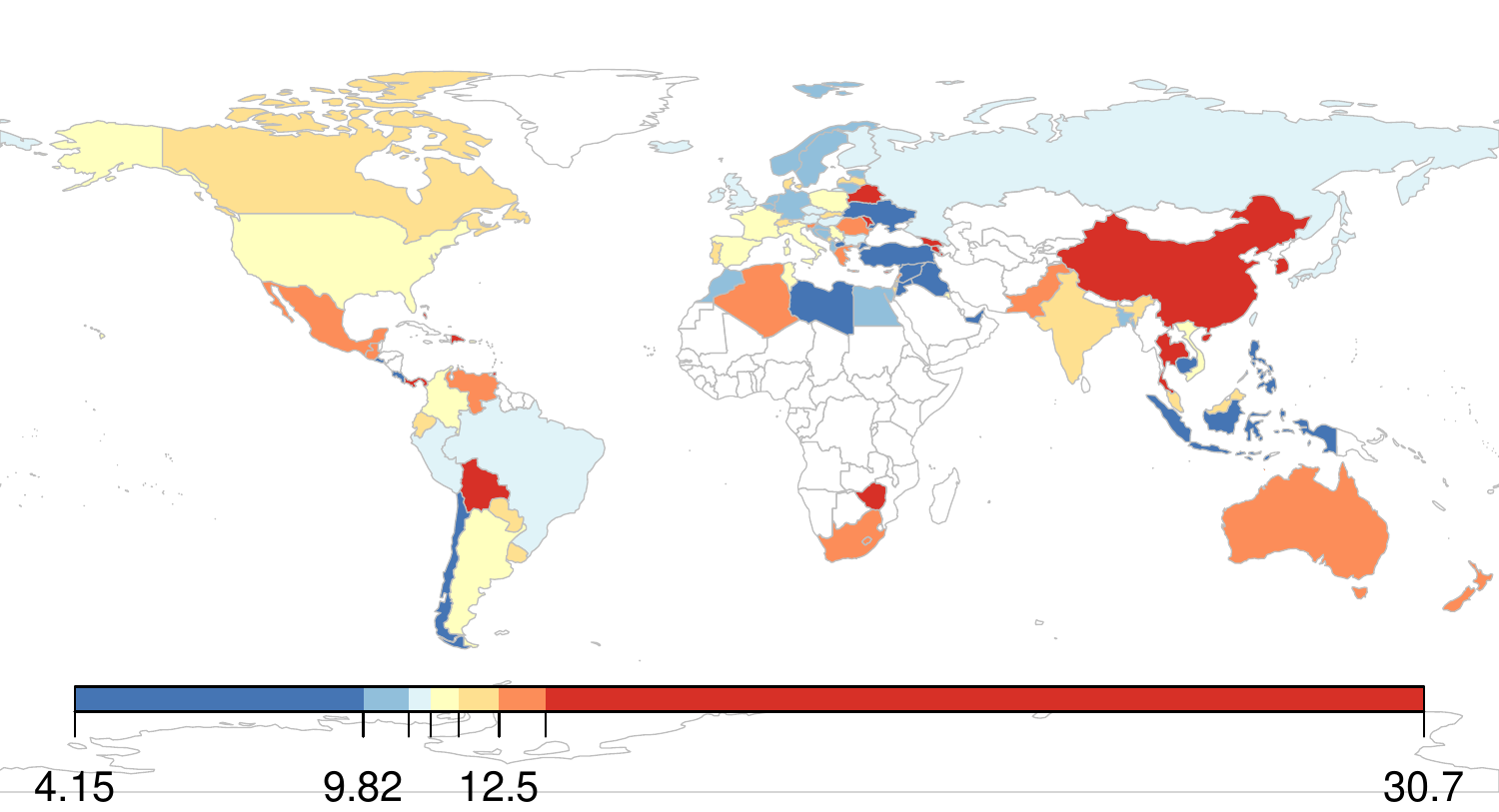}
 	\caption{Heat map showing the percentage of posts with hate speech per country. [Best viewed in color.]}
 	\label{fig:hate-map}
	\vspace{-0.5cm}
 \end{figure}

Figure~\ref{fig:hate-map} plots a heat map of the percentage of posts that contain hate speech per country with at least 1,000 posts on \dspol.
Countries are placed into seven equally populated bins and colored from blue to red depending on the percentage of their posts contain a hate word from the hatebase dictionary.

Note that some of the most ``hateful'' countries (e.g., Bahamas and Zimbabwe) might be overrepresented due to the use of proxies in those countries.
Zimbabwe is of particular interest to \dspol users because of its history as the unrecognized state of Rhodesia.

To understand whether the country flag has any meaning, we run a term frequency-inverse document frequency (TF-IDF) analysis to identify topics that are used per country.
We remove all countries that have less than 1,000 posts, as this eliminates the most obvious potential proxy locations.
After removing stop words and performing stemming, we build TF-IDF vectors for each of the remaining 98 countries, representing the frequencies with which different words are used, but down-weighted by the general frequency of each word across all countries.
When examining the TF-IDF vectors, although we cannot definitively exclude the presence of proxied users, we see that the majority of posts from countries seem to match geographically, e.g., posters from the US talk about Trump and the elections more than posters from South America, users in the UK talk about Brexit, those from Greece about the economic and immigration crisis, and people from Turkey about the attempted coup in July 2016.

\descr{Clustering.} To provide more evidence for the conclusion that \dspol is geo-politically diverse, we perform some basic text classification and evaluate whether or not different parts
of the world are talking about ``similar'' topics.
We apply spectral clustering over the vectors using the Eigengap heuristic~\cite{ng2002spectral} to automatically identify the number of target clusters.
In Figure~\ref{fig:country-cluster-map}, we present a world map colored according to the 8 clusters generated.
Indeed, we see the formation of geo-political ``blocks.''
Most of Western Europe is clustered together, and so are USA and Canada, while the Balkans are in a cluster with Russia.
One possible limitation stemming from our spectral clustering is its sensitivity to the total number of countries we are attempting to cluster.
Indeed, we find that, by filtering out fewer countries based on number of posts, the clusters \emph{do} change.
For instance, if we do not filter any country out, France is clustered with former French colonies and territories, Spain with South America, and a few of the Nordic countries flip between the Western Europe and the North American clusters.
Additionally, while \dspol posts are almost exclusively in English, certain phrasings, misspellings, etc. from non native speakers might also influence the clustering.
That said, the overall picture remains consistent: the flags associated with \dspol posts are meaningful in terms of the topics those posts talk about.

\section{Raids Against Other Services}\label{sec:raids}
As discussed previously,
\dspol is
often used to post links to other sites:
some are posted to initiate discussion or provide additional commentary,
but others serve to call \dspol users to certain coordinated actions, including
attempts to skew post-debate polls~\cite{dailydot} as well as ``raids''~\cite{tumblr}.

Broadly speaking, a raid is an attempt to disrupt another site,
not from a network perspective (as in a DDoS attack), but from
a content point of view.
I.e., raids are not an attempt to directly attack a 3rd party service itself, but rather to disrupt the \emph{community} that calls that service home.
Raids on \dspol are semi-organized: we anecdotally observe a number of calls for action~\cite{bernstein20114chan}
consisting of a link to a target -- e.g., a YouTube video or a Twitter hashtag -- and the text ``you know what to do,''
prompting other 4chan users to start harassing the target.
The thread itself often becomes an aggregation point with screenshots of the target's reaction, sharing of sock puppet accounts used to harass, etc.

In this section, we study how raids on YouTube work.
We show that synchronization between \dspol threads and YouTube comments is correlated with an increase in hate speech in the YouTube comments.
We further show evidence that the synchronization is correlated with a high degree of overlap in YouTube commenters. First, however, we discuss a case study of a very broad-target raid, attempting to mess
with anti-trolling tools by substituting racially charged words with company names, e.g., ``googles.''

\subsection{Case Study: ``Operation Google''}\label{sec:op-google}
We now present with a case study of a very broad-target raid, attempting to mess
with anti-trolling tools by substituting racially charged words with company names, e.g., ``googles.''
On September 22, 2016, a thread on \dspol called for the execution of so-called ``Operation Google,'' in response to Google announcing the introduction of anti-trolling machine learning based technology~\cite{wired2} and similar initiatives on Twitter~\cite{huffpo}.
It was proposed to poison these by using, e.g., ``Google'' instead of ``nigger'' and ``Skype'' for ``kike,'' calling other users to disrupt social media sites like Twitter, and also recommending using certain hashtags, e.g., \hasht{worthlessgoogs} and \hasht{googlehangout}.
By examining the impact of Operation Google on both \dspol and Twitter, we aim to gain useful insight into just how efficient and effective the \dspol community is in acting in a coordinated manner.

\begin{figure}[t]
    \centering
    \includegraphics[width=0.85\columnwidth]{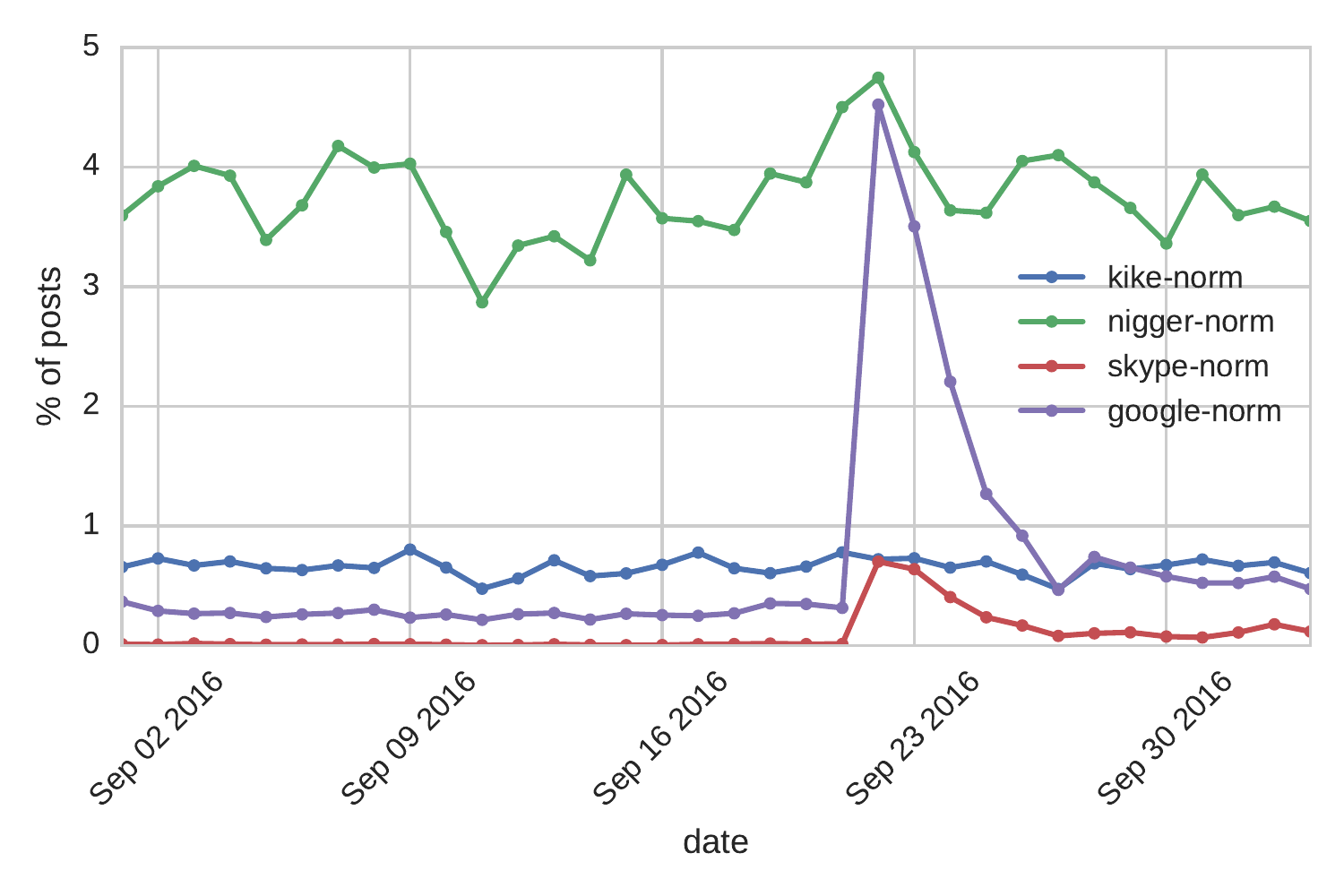}
    \vspace{-0.3cm}
    \caption{The effects of Operation Google within \dspolCCC.}
    \label{fig:op-google-pol}
    \vspace{-0.2cm}
\end{figure}

In Figure~\ref{fig:op-google-pol}, we plot the normalized usage of the specific replacements called for in the Operation Google post.
The effects within \dspol are quite evident: on Sep 22 we see the word ``google'' appearing at over 5 times its normal rate, while ``Skype'' appears at almost double its normal rate.
To some extent, this illustrates how quickly \dspol can execute on a raid, but also how short of an attention span its users have: by Sep 26 the burst in usage of Google and Skype had died down.
While we still see elevated usages of ``Google'' and ``Skype,'' there is no discernible change in the usage of ``nigger'' or ``kike,''
but these replacement words do seem to have become part of \dspol's vernacular.

\begin{figure}[t]
    \centering
    \includegraphics[width=0.85\columnwidth]{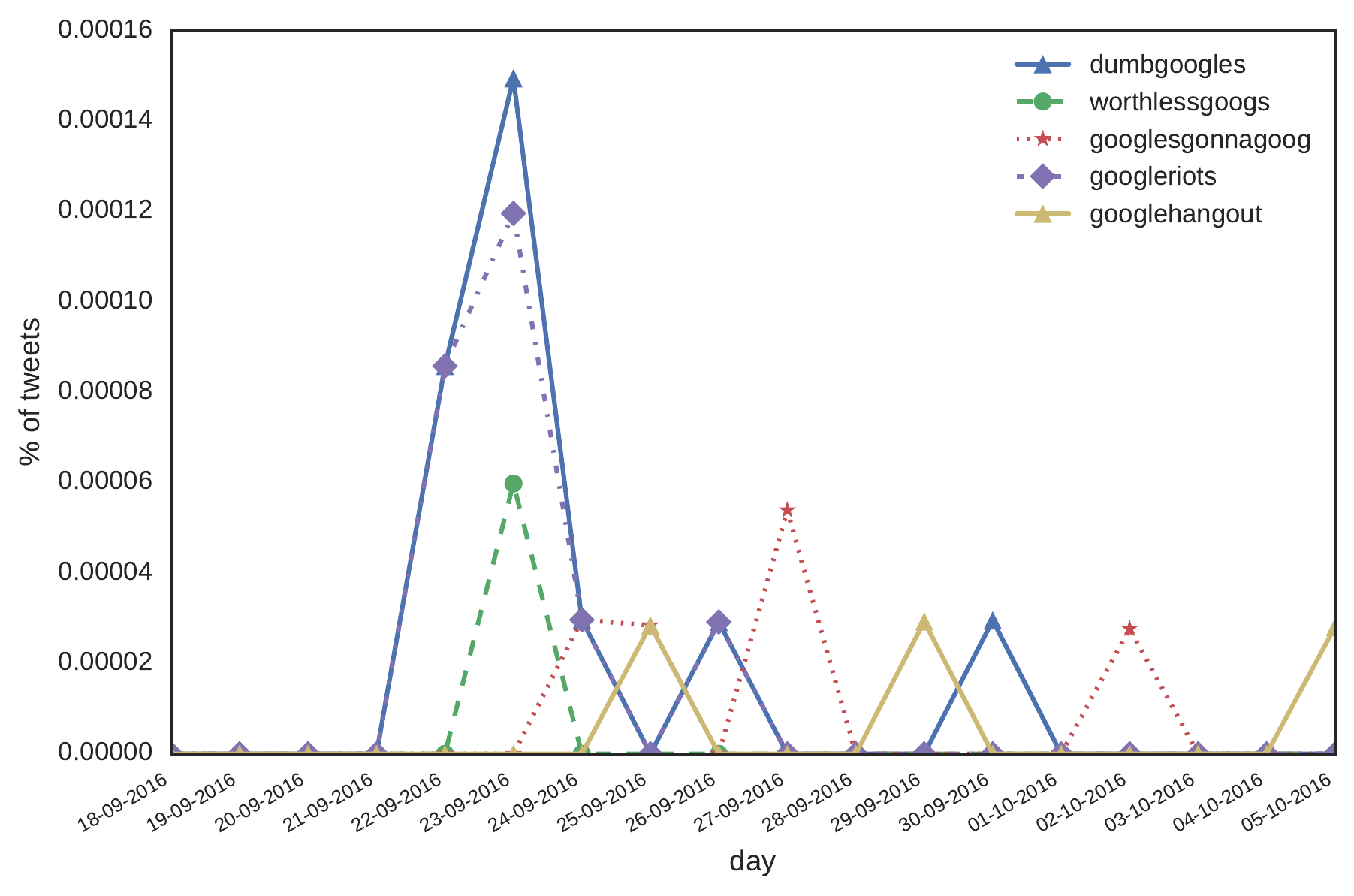}
    \vspace{-0.3cm}
    \caption{The effects of ``Operation Google'' on Twitter.}
    \label{fig:op-google-twitter}
\end{figure}

\begin{figure}[t]
\centering
    \begin{subfigure}[b]{0.37\textwidth}
    \includegraphics[width=1\linewidth]{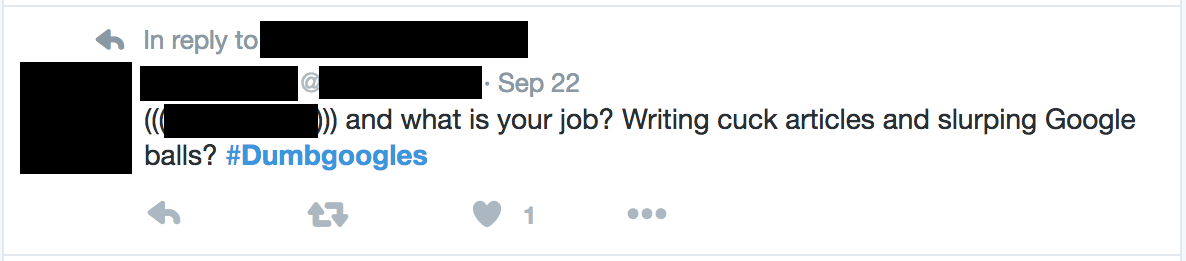}
    \caption{}
\end{subfigure}

\begin{subfigure}[b]{0.37\textwidth}
    \includegraphics[width=1\linewidth]{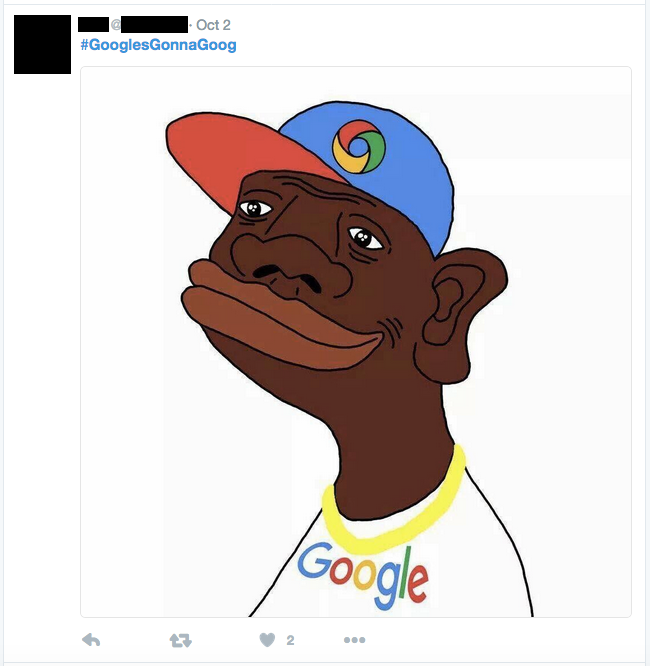}
    \caption{}
\end{subfigure}
\caption[]{Two tweets featuring Operation Google hashtags in combination with other racist memes. }
\label{fig:op-google-tweets}
\end{figure}

Next, we investigate the effects of Operation Google \emph{outside} of \dspol,
counting how many tweets in our 60M tweet dataset (see Section~\ref{sec:dataset}) %
contain the hashtags \hasht{worthlessgoogs}, \hasht{googlehangout}, \hasht{googleriots}, \hasht{googlesgonnagoog}, and \hasht{dumbgoogles} in Figure~\ref{fig:op-google-twitter}. (Recall that our dataset consists of a 1\% sample of all public
tweets from Sep 18 to Oct 5, 2016.)
Figure~\ref{fig:op-google-tweets} provides two example tweets from our dataset that contain Operation Google hashtags.
As expected, the first instances of those hashtags, specifically, \hasht{googleriots} and \hasht{dumbgoogles}, appear on Sep 22. On Sep 23, we also see \hasht{worthlessgoogs} and, on later days, the rest of the hashtags. Overall, Sep 23 features the highest hashtag activity during our observation period.
While this does indicate an attempt to instigate censorship evasion on Twitter, the percentage of tweets containing these hashtags shows that Operation Google's impact was much more prevalent on \dspol itself than on Twitter. For example, on Sep. 23, \hasht{dumbgoogles} appears in only 5 out 3M tweets (0.00016\%) in our dataset for that day, despite it being the most ``popular'' hashtag (among the ones involved in Operation Google) on the most ``active'' day.
Incidentally, this is somewhat at odds with the level of media coverage around Operation Google~\cite{telegraph}.

\subsection{Spreading Hate on YouTube}

As discussed in our literature review,
we still have limited insight into how trolls operate,
and in particular how forces outside the control of targeted services organize and coordinate their actions.
To this end, we set out to investigate the connection between \dspol threads and YouTube comments. We focus on YouTube
since 1)~it accounts for the majority of media links posted on \dspol,
and 2)~it is experiencing an increase in hateful comments, prompting Google to announce the (not uncontroversial) YouTube Heroes program~\cite{heroes}.

We examine the comments from 19,568 YouTube videos linked to by 10,809 \dspol threads to look for raiding behavior at scale.
Note that finding evidence of raids on YouTube (or any other service)
is not an easy task, considering that explicit calls for raids are an offense that
can get users banned.\footnote{Recall that, since there are no accounts on 4chan, bans are based on session/cookies or IP addresses/ranges, with the latter causing VPN/proxies to be banned often.}
Therefore, rather than looking for a particular trigger on \dspol, we look for elevated activity in comments on YouTube videos linked from \dspol. In a nutshell, we expect raids to exhibit synchronized activity between comments in a \dspol thread a YouTube link appears in and the amount of comments it receives on YouTube. We also expect the rate of hateful comments to increase after a link is posted on \dspol.

\subsection{Activity Modeling}
To model synchronized activities, we use
signal processing techniques. First, we introduce some notation: Let
$x$ be a \dspol thread, and $y$ the set of comments to a YouTube
video linked from $x$. We denote with
$\left\{t_x^i|i=1,..N_x\right\}$ and
$\left\{t_y^j|j=1,..N_y\right\}$, respectively, the set of
timestamps of posts in $x$ and $y$. Since the lifetime of \dspol
threads is quite dynamic, we shift and normalize the time axis for
both $\left\{t_x^i\right\}$ and $\left\{t_y^j\right\}$, so that
$t=0$ corresponds to when the video was first linked and $t=1$ to
the last post in the \dspol thread:

\vspace*{-0.5\baselineskip}
{\small\begin{equation*}\label{eq:time}
    t \leftarrow \frac{t-t_{yt}}{t_{last}-t_{yt}} .
\end{equation*}}
\vspace*{-\baselineskip}
\\[-6pt]

In other words, we normalize to the duration of the \dspol thread's
lifetime. We consider only \dspol posts that occur after the YouTube
mention, while, for computational complexity reasons, we consider
only YouTube comments that occurred within the (normalized) $[-10,
+10]$ period, which accounts for $~35\%$ of YouTube comments in our
dataset.

\begin{figure}[t]
    \centering
    \includegraphics[width=0.8\columnwidth]{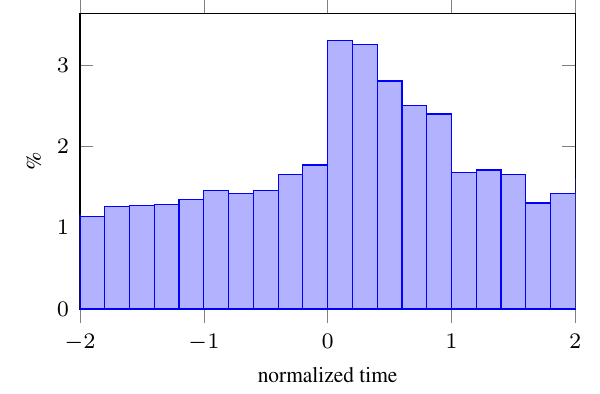}
        \vspace{-0.2cm} \caption{Distribution of the distance (in normalized
        thread lifetime) of the highest peak of activity in YouTube comments
        and the \dspolCCC thread they appear in. $t = 0$ denotes the time when
        video was first mentioned, and $t = 1$ the last related post in the
        thread.} \label{fig:peaks_histogram} \vspace{-0.05cm}
\end{figure}

From the list of YouTube comment timestamps, we compute the
corresponding Probability Density Function (PDF) using the Kernel
Density Estimator method~\cite{silverman1986density}, and estimate the position of the
absolute maximum of the distribution.
In Figure~\ref{fig:peaks_histogram}, we plot the distribution of the
distance between the highest peak in YouTube commenting activity and
the \dspol post linking to the video. We observe that 14\%
of the YouTube videos experience a peak in activity during
the period they are discussed on \dspol. In many cases,
\dspol seems to have a strong influence on YouTube activity,
suggesting that the YouTube link posted on \dspol might have a triggering behavior,
even though this analysis does not necessarily provide evidence of a raid taking place.

However, if a raid \emph{is} taking place, then the comments on both
\dspol and YouTube are likely to be ``synchronized.''
Consider, for instance, the extreme case where some users that see the YouTube
link on a \dspol thread comment on both YouTube and and the \dspol thread
simultaneously: the two set of timestamps would be perfectly synchronized.
In practice, we measure the synchronization, in terms of delay
between activities,
using {\em cross-correlation} to estimate the lag between two signals.
In practice, cross-correlation slides one signal with respect to the
other and calculates the dot product (i.e., the \emph{matching})
between the two signals for each possible lag. The estimated lag is
the one that maximizes the matching between the signals.
We represent the sequences as signals ($x(t)$ and $y(t)$), using
Dirac delta distributions $\delta(\cdot)$. Specifically,
we expand $x(t)$ and $y(t)$ into trains of Dirac delta
distributions:

\vspace*{-0.75\baselineskip}
{\small\begin{equation*}\label{eq:dirac_trains}
    x(t) = \sum_{i=1}^{N_x} \delta\left(t-t_x^i\right) ; \;\;
    y(t) = \sum_{j=1}^{N_y} \delta\left(t-t_y^j\right)
\end{equation*}}
\vspace*{-0.75\baselineskip}

\noindent
and we calculate $c(t)$, the continuous time cross-correlation
between the two series\footnote{Since timestamp resolution is 1s,
this is equivalent to a discrete-time cross-correlation with 1s
binning, but the closed form solution lets us compute it much more
efficiently.} as:
\vspace*{-0.5\baselineskip}
{\small\begin{align*}\label{eq:trains_correlation}
        c(t) = \int_{-\infty }^{\infty}x(t+\tau)y(\tau)d\tau = \sum_{i=1}^{N_x} \sum_{j=1}^{N_y} \delta\left(t-\left(t_y^j-t_x^i\right)\right)
\end{align*}}
\vspace*{-\baselineskip}

The resulting cross-correlation is also a Dirac delta train,
representing the set of all possible inter-arrival times between
elements from the two sets.

If $y(t)$ is the version of $x(t)$ shifted by
 $\Delta T$ (or at least contains a shifted version of
$x(t)$), with each sample delayed with a slightly different time
lag, $c(t)$ will be characterized by a high concentration of pulses
around $\Delta T$. As in the peak activity detection, we can
estimate the more likely lag by computing the associated PDF
function $\hat{c}(t)$ by means of the Kernel Density Estimator
method~\cite{silverman1986density}, and then compute the global
maximum:

\vspace*{-0.5\baselineskip}
{\small
\begin{equation*}\label{eq:smoothed_correlation}
\hat{c}(t) = \int_{-\infty }^{\infty}c(t+\tau)k(\tau)d\tau  ; \; \; \hat{\Delta T} = \arg\max_t \hat{c}(t)
\end{equation*}
}
\vspace*{-0.75\baselineskip}

\noindent where $k(t)$ is the kernel smoothing function (typically a
zero-mean Gaussian function).\footnote{$\hat{c}(t)$ is also the
cross-correlation between the PDF functions related to $x(t)$ and
$y(t)$.}

\subsection{Evidence of Raids}

\begin{figure}[t]
    \centering
    \includegraphics[width=0.8\columnwidth]{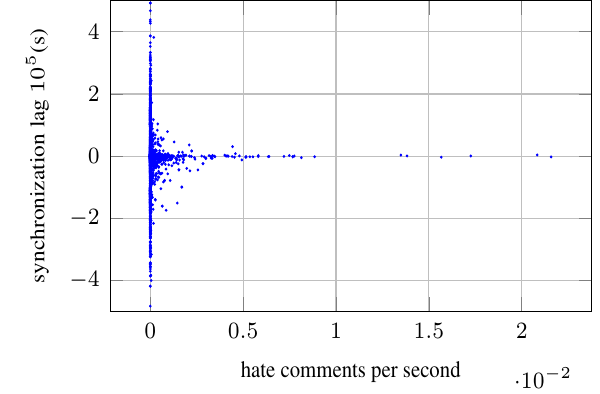}
        \vspace{-0.2cm} \caption{Hateful YouTube comments vs synchronization
        lag between \dspolCCC threads and corresponding YouTube comments.
        Each point is a \dspolCCC thread. The hateful comments count refers to just those within the thread lifetime ([0,+1])}
        \label{fig:hate-vs-lag}
        \vspace{-0.05cm}
\end{figure}

Building on the above insights, we provide large-scale evidence of raids.
If a raid is taking place, we expect the estimated lag $\Delta T$ to
be close to zero, and we can validate this by looking at the content
of the YouTube comments.

Figure~\ref{fig:hate-vs-lag} plots the
relationship between the number of {\em hateful} comments on YouTube
that occur within the \dspol thread lifetime (i.e., containing at
least one word from the hatebase dictionary) and the synchronization
lag between the \dspol thread and the YouTube comments. The trend is
quite clear: as the rate of hateful comments on YouTube increases,
the synchronization lag between \dspol and YouTube comments
decreases. This  shows that almost all YouTube videos affected
by (detected) hateful comments during the \dspol thread lifetime are
likely related to raids.

\begin{figure}[t]
    \centering
    \includegraphics[width=0.75\columnwidth]{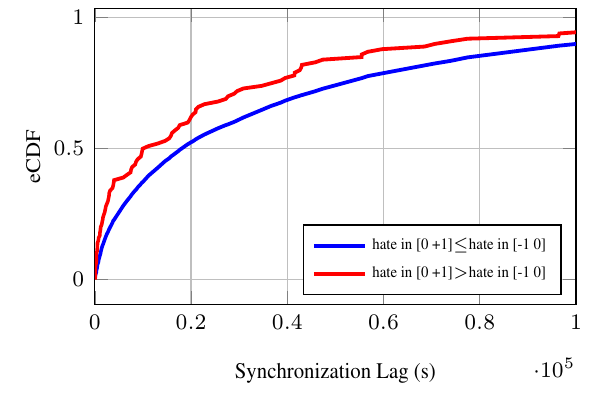}
        \vspace{-0.2cm}
        \caption{CDF of synchronization lag between
        \dspolCCC threads and YouTube comments, distinguishing between
        threads with YouTube videos containing higher hate comments percentage in the [0
        +1] period or [-1 0].}
        \label{fig:ks-lag-within-before-hate}
        \vspace{-0.05cm}
\end{figure}

Figure~\ref{fig:ks-lag-within-before-hate} plots the CDF of the absolute
value of the synchronization lag between \dspol threads and comments
on the corresponding YouTube videos. We distinguish between comments
with a higher percentage of comments containing hate words {\em
during} the life of the thread from those with more {\em before} the
thread. In other words, we compare threads where \dspol appears to
have a negative impact vs. those where they do not. From the plot,
we observe that the YouTube comments with more hate speech during
the \dspol thread's lifetime are significantly ($p < 0.01$ with a 2-sample Kolmogorov-Smirnov test) more synchronized
with the \dspol thread itself.

Finally, to further show that \dspol
is raiding YouTube videos, we can look at the authors of YouTube
comments.
We argue that, unlike the anonymous venue of \dspol, raids on a service like YouTube will leave evidence via account usage, and that the same raiding YouTube accounts will likely be used by \dspol users more than once. Indeed, while it is moderately easy to create a new YouTube account, there is still some effort involved. Troll accounts might also be cultivated for use over time, gaining some reputation as they go along.
Perhaps more importantly, while less anonymous than \dspol, YouTube accounts are still only identified by a profile name and do not truly reveal the identity of the user.

\begin{figure}[t]
    \centering
    \includegraphics[width=0.75\columnwidth]{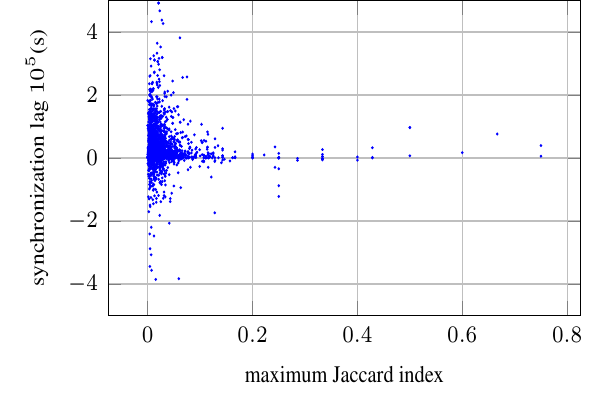}
        \vspace{-0.2cm} \caption{Maximum Jaccard Index of a YouTube video and all others vs synchronization
        lag between \dspolCCC threads and corresponding YouTube comments.
        Note the high correlation between overlap and synchronization lag.}
        \label{fig:overlap_vs_sync}
        \vspace{0.0cm}
\end{figure}

To measure this,
we compute the overlap (Jaccard index) of commenters in each YouTube video.
In Figure~\ref{fig:overlap_vs_sync} we plot the
synchronization lag as a function of the maximum overlap
between a given video and all others.
From the figure we observe that if a YouTube video has relatively high overlap with at least one other YouTube video, it also highly synchronized with its corresponding \dspol thread, indicative of a raid taking place.

\section{Discussion \& Conclusion}\label{sec:conclusion}
This paper presented the first large-scale study of \dspol, 4chan's politically incorrect board, arguably the most controversial one owing to its links to the alt-right movement and its
unconventional support to Donald Trump's 2016 presidential campaign.
First, we provided a general characterization, comparing activity on \dspol to two other boards on 4chan, \dssp (``sports'') and \dsint (``international'').
We showed that each of the boards exhibits different behaviors with respect to thread creation and posts. We looked at the impact of ``bump limits'' on discourse, finding that it results in fresh content on a consistent basis.
We used the country flag feature present on the three boards and found that, while Americans dominate the conversation in terms of absolute numbers, many other countries (both native English speaking and not) are well represented in terms of posts per capita.
We also showed differences in the maturity of threads with respect to moderators' actions across the boards.

Next, we examined the content posted to \dspol, finding that the majority of links posted to the board point to YouTube.
We also saw that \dspol contains many more links to tabloid and right-wing leaning news outlets than mainstream sites.
By looking at metadata associated with posted images, we learned that most content on 4chan is quite unique: 70\% of the 1M unique images in our dataset were posted only once and 95\% less than 5 times.
In fact, \dspol's ability to find or produce original content is likely one of the reasons it is thought to be at the center of hate on the web.

Finally, we studied ``raiding'' behavior
by looking for evidence of \dspol's hateful impact on YouTube comments.
We used signal processing techniques to discover that peaks of commenting activity on YouTube tend to occur within the lifetime of the thread they were posted to on \dspol.
Next, we used cross-correlation to estimate the synchronization lag between \dspol threads and comments on linked YouTube videos.
Here, we found that as the synchronization lag approaches zero, there is an increase in the rate of comments with hate words on the linked YouTube comments.
Finally, we saw that if two YouTube videos' comments had many common authors they were likely to be highly synchronized, indicating potential raider accounts.
This evidence suggests that, while not necessarily explicitly called for (and in fact, against \dspol's rules), \dspol users \emph{are} performing raids in an attempt to disrupt the community of YouTube users.

Overall, our analysis provides not only the first measurement study of \dspol, but also insight into the continued growth of hate and extremism trends on social media, and prompts a few interesting problems for future research.
Naturally, however, our work is not without limitations.
First, although the Hatebase dataset we used is an invaluable resource for hate speech analysis,
the usage of ``hate'' words may be context-dependent, and we leave it to future work to investigate how to distinguish context (e.g., by recognizing sarcasm or trolling).
Also, our flag based country analysis may have been influenced by the use of VPNs/proxies:
although this does not affect the validity of our results, it calls for a more in-depth analysis of language and posting behavior.
Finally, while we showed quantitative evidence that raids are taking place,
we do not claim an ability to \emph{classify} them as there are many layers of subtlety in how raiding behavior might be exhibited. However, we are confident that our findings can serve as a foundation for interesting and valuable future work exploring fringe groups like the alt-right, hate speech, and online harassment campaigns.

\descr{Acknowledgments.} We wish to thank Andri Ioannou and Despoina
Chatzakou for their help and feedback, and Timothy Quinn for providing access to the Hatebase API.
This research is supported
by the European Union's H2020-MSCA-RISE grant ``ENCASE''
(GA No. 691025) and by the EPSRC under
grant EP/N008448/1. Jeremiah Onaolapo was supported by the
Petroleum Technology Development Fund (PTDF).

\small
\bibliographystyle{abbrv}
\bibliography{bibfile}

\section*{Appendix}
\appendix

\section{Rare Pepes}\label{sec:pepes}

In this Section we display some of our rare Pepe collection.

\begin{figure}[h]
	\centering
	\includegraphics[width=\columnwidth]{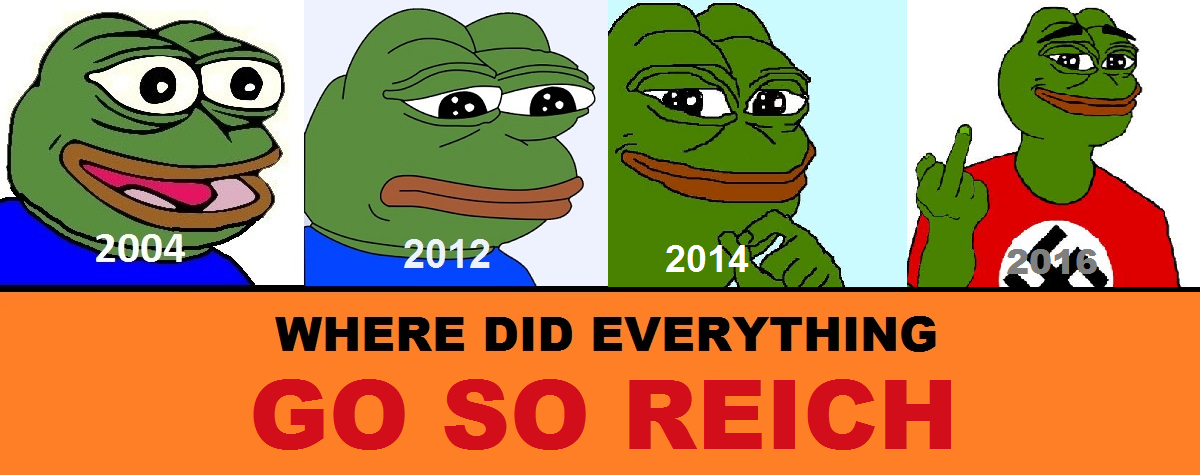}
	\caption{A somewhat rare, modern Pepe, which much like the Bayeux Tapestry records the historic rise of \dspol.}
	\label{fig:pepe1}
\end{figure}

\begin{figure}[h]
	\centering
	\includegraphics[width=\columnwidth]{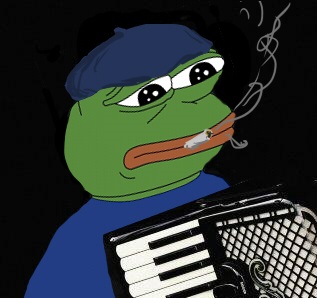}
	\caption{A (French?) Pepe wearing a beret, smoking a cigarette, and playing an accordion.}
	\label{fig:pepe3}
\end{figure}

\begin{figure}[h]
	\centering
	\includegraphics[width=0.8\columnwidth]{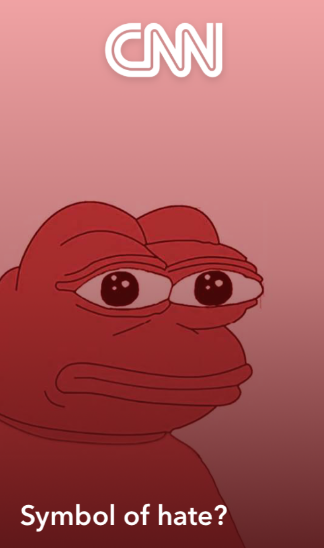}
	\caption{An extremely common Pepe commissioned by CNN to commemorate Pepe's recognition as a hate symbol.}
	\label{fig:pepe2}
\end{figure}

\begin{figure}[h]
	\centering
	\includegraphics[width=\columnwidth]{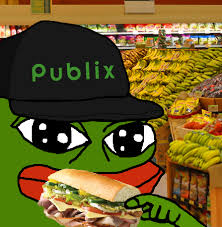}
	\caption{An (unfortunately) ultra rare Pepe eating a delicious Publix Deli Sub Sandwich.}
	\label{fig:pepe4}
\end{figure}

\begin{figure}[h]
	\centering
	\includegraphics[width=\columnwidth]{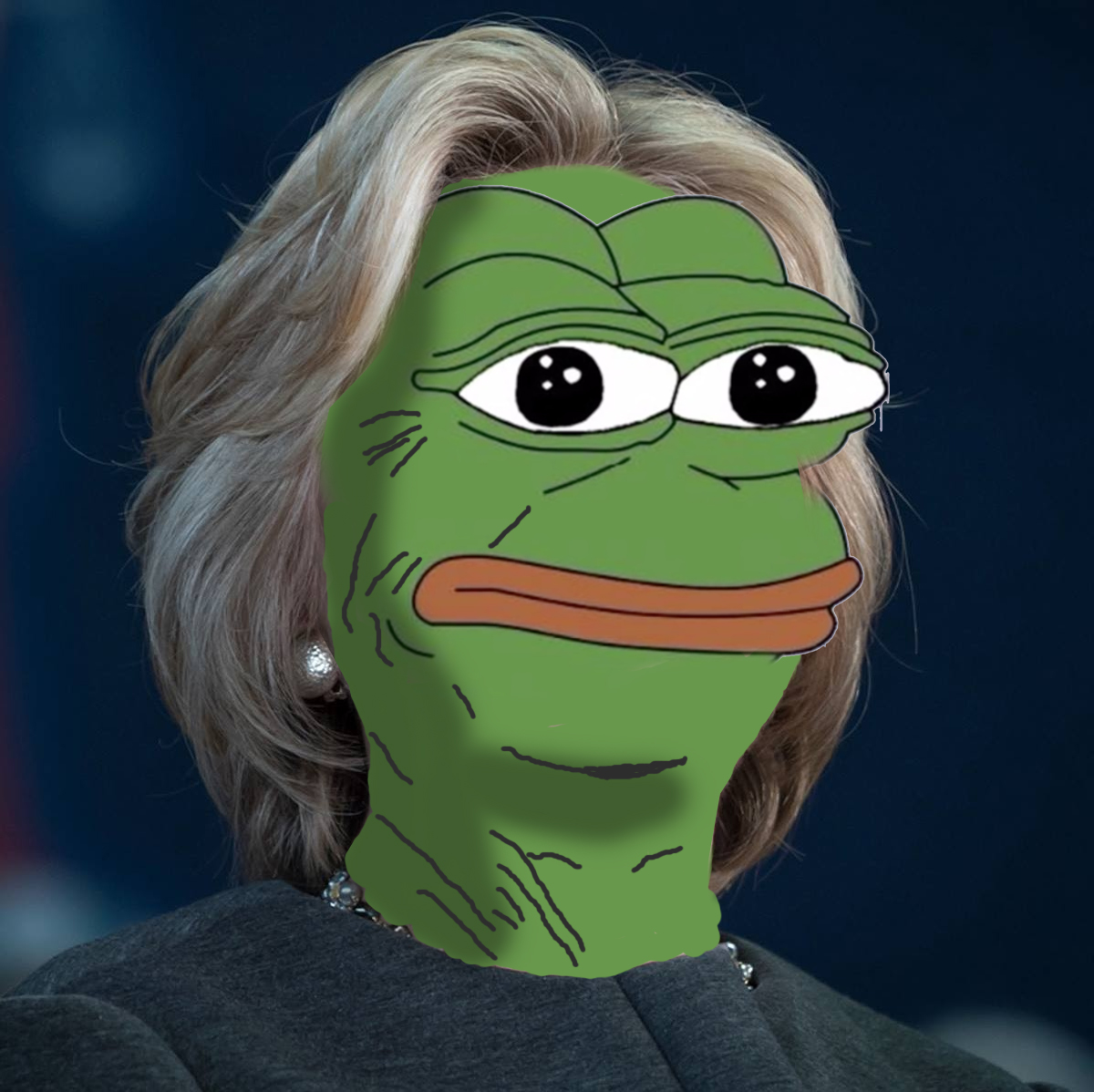}
	\caption{An ironic Pepe depiction of Hillary Clinton.}
	\label{fig:pepe5}
\end{figure}

\begin{figure}[h]
	\centering
	\includegraphics[width=\columnwidth]{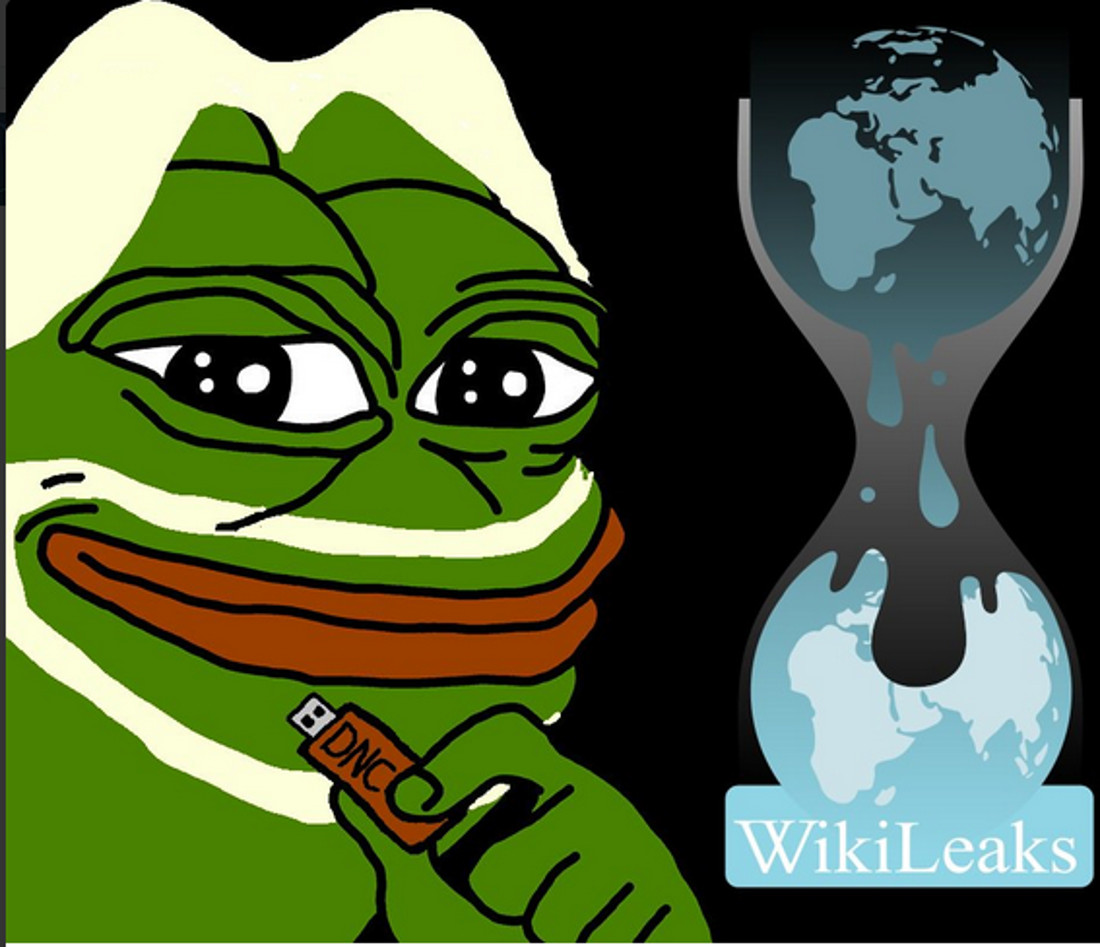}
	\caption{A Pepe Julian Asange dangling a USB full of Democratic National Convention secrets.}
	\label{fig:pepe6}
\end{figure}

\begin{figure}[h]
	\centering
	\includegraphics[width=\columnwidth]{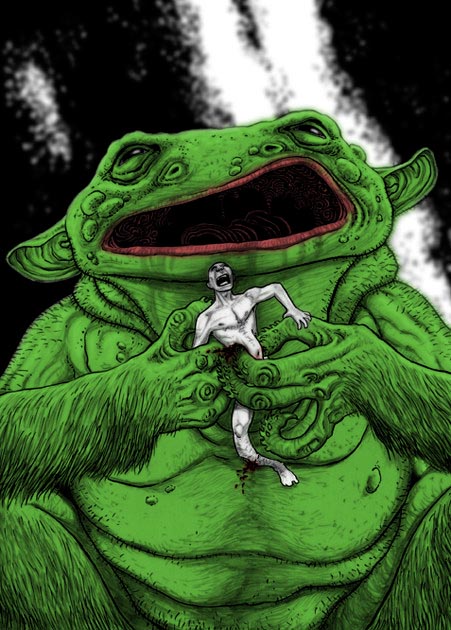}
	\caption{What we believe to be a Pepe re-interpretation of Goya's ``Saturn Devouring His Son.''}
	\label{fig:pepe7}
\end{figure}

\begin{figure}[h]
	\centering
	\includegraphics[width=\columnwidth]{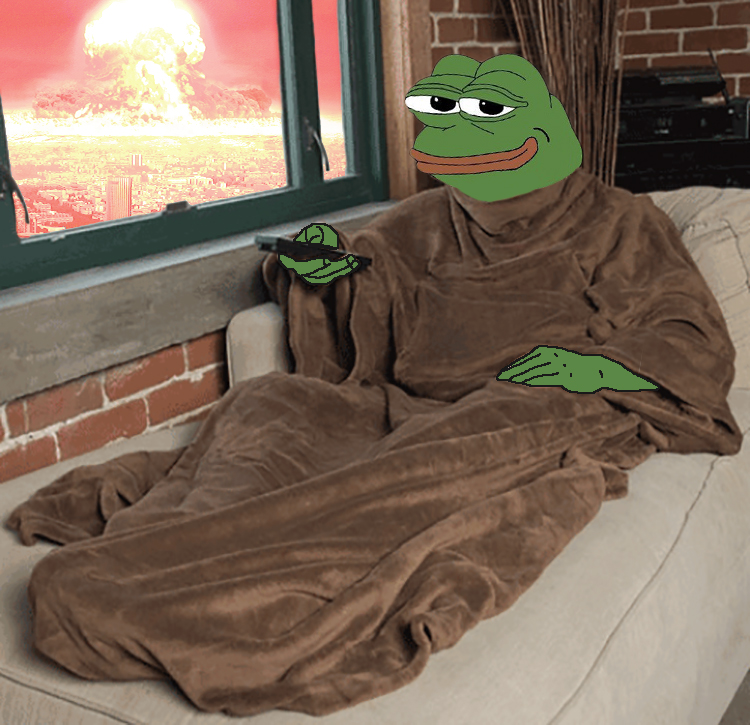}
	\caption{A very comfy Pepe.}
	\label{fig:pepe8}
\end{figure}

\begin{figure*}[p]
    \centering
    \begin{minipage}[t]{0.475\textwidth}
        \centering
	\includegraphics[width=\columnwidth]{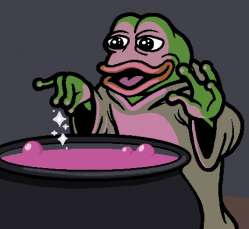}
	\caption{A mischievous witch Pepe.}
	\label{fig:pepe9}
	\end{minipage}
	\hfill
    \begin{minipage}[t]{0.475\textwidth}
	\centering
	\includegraphics[width=\columnwidth]{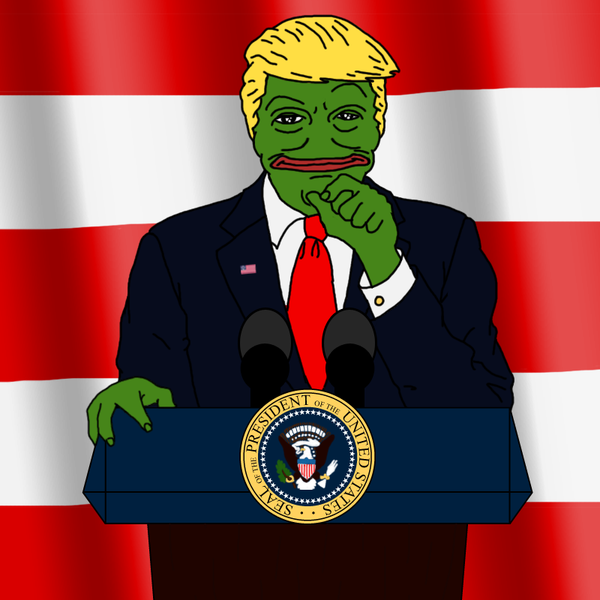}
	\caption{The now ``iconic'' Trump Pepe.}
	\label{fig:pepe10}
		\end{minipage}

\end{figure*}

\end{document}